\documentclass[useAMS,usenatbib,referee]{biom}

\usepackage{amsmath,amssymb,wasysym}
\usepackage{color}
\usepackage[figuresright]{rotating}

\def\bSig\mathbf{\Sigma}

\title[Multiple Testing for Neuroimaging via Hidden Markov Random Field]{Multiple Testing for Neuroimaging via Hidden Markov Random Field}

\author{Hai Shu$^{1,*}$\email{haishu@umich.edu},
	Bin Nan$^{1,**}$\email{bnan@umich.edu},
	Robert Koeppe$^{2,***}$\email{koeppe@umich.edu} \\
	$^{1}$Department of Biostatistics, University of Michigan, Ann Arbor, Michigan, U.S.A.\\
	$^{2}$Department of Radiology, University of Michigan, Ann Arbor, Michigan, U.S.A.}

\begin{document}





\pagerange{\pageref{firstpage}--\pageref{lastpage}}
\volume{000}
\pubyear{0000}
\artmonth{000}


\doi{000}


\label{firstpage}


\begin{abstract}
Traditional voxel-level multiple testing procedures in neuroimaging, mostly $p$-value based, often ignore the spatial correlations among neighboring voxels and thus suffer from substantial loss of power. We extend the local-significance-index based procedure originally developed for the hidden Markov chain models, which aims to minimize the false nondiscovery rate subject to a constraint on the false discovery rate, to three-dimensional neuroimaging data using a hidden Markov random field model. A generalized expectation-maximization algorithm
for maximizing the penalized likelihood
is proposed for estimating the model parameters. Extensive simulations show that the proposed approach is more powerful than conventional false discovery rate procedures. We apply the method to the comparison between mild cognitive impairment, a disease status with increased risk of developing Alzheimer's or another dementia, and normal controls in the FDG-PET imaging study of the Alzheimer's Disease Neuroimaging Initiative. \\
\end{abstract}

%


\footnotetext{This is the peer reviewed version of the following article: [Shu, H., Nan, B., and Koeppe, R. (2015).
	Multiple testing for neuroimaging via hidden Markov random field. Biometrics, 71(3), 741-750.],
	which has been published in final form at [DOI: 10.1111/biom.12329].
	This article may be used for non-commercial purposes in accordance with
	Wiley Terms and Conditions for Self-Archiving.}


\maketitle


%

\section{Introduction}
\label{s:intro}
In a seminal paper, \citet{Benj95} introduced false discovery rate (FDR) as an alternative measure of Type I error in multiple testing problems to the family-wise error rate (FWER).
They showed that the FDR is equivalent to the FWER if all null hypotheses are true and is smaller otherwise, thus FDR controlling procedures potentially have a gain in power over FWER controlling procedures.
FDR is defined as the expected proportion of false rejections among all rejections. The false nondiscovery rate
(FNR; Genovese and Wasserman, 2002), the expected proportion of falsely accepted hypotheses, is the corresponding measure of Type II error.
The traditional FDR procedures \citep{Benj95, Benj00,Geno04},
which are $p$-value based, are theoretically developed under the  assumption that the test statistics are independent.
Although these approaches are shown to be valid in controlling FDR under certain dependence assumptions \citep{Benj01,Farc07,Wu08}, they may
suffer from severe loss of power when the dependence structure is ignored
\citep{Sun09}.
By modeling the dependence structure using a hidden Markov chain (HMC),
\citet{Sun09} proposed
an oracle FDR procedure built on a new test statistic, the local index of significance (LIS),  and the corresponding asymptotic data-driven procedure, which are optimal in the sense that they minimize the marginal FNR subject to a constraint on the marginal FDR.
Following the work of
\citet{Sun09}, \citet{Wei09} developed a pooled LIS (PLIS) procedure for multiple-group analysis where different groups have different HMC dependence structures, and proved the optimality of the PLIS procedure.
Either the LIS procedure or the PLIS procedure only handles the one-dimensional dependency.
However, problems with higher dimensional dependence are of particular practical interest in analyzing imaging data.

FDR procedures have been widely used in analyzing neuroimaging data, such as positron emission tomography (PET) imaging and functional magnetic resonance imaging (fMRI) data (Genovese, Lazar, and Nichols, 2002; Chumbley and Friston, 2009; Chumbley et al., 2010, among many others).
We  extend the work of \citet{Sun09}
 in this article
by developing
 an optimal LIS-based FDR procedure for three-dimensional (3D) imaging data using a hidden Markov random field model (HMRF) for the spatial dependency among multiple tests. Existing methods for correlated imaging data, for example, \citet{Zhan11} are not  shown to be optimal,
i.e., minimizing FNR.

HMRF model is a generalization of HMC model, which replaces the underlying Markov chain by Markov random field.
A well-known classical Markov random field with two states is the Ising model.
In particular, the two-parameter Ising model,
 whose formal definition is given in Equation (\ref{Ising model}),
reduces to the two-state Markov chain in one-dimension \citep{Brem99}. The Ising model and its generalization with more than two states, the Potts model, have been widely used to capture the spatial structure in image analysis; see \citet{Brem99}, \citet{ Wink03}, \citet{Zhan08}, \citet{Huan13} and \citet{John13}, among others.
In this article, we consider a hidden Ising model for each area based on
the Brodmann's partition of the cerebral cortex \citep{Gare06} and subcortical regions of the human brain, which provides a natural way of modeling spatial correlations for neuroimaging data. To the best of our knowledge, this is the first work that introduces the HMRF-LIS based FDR procedure to the field of neuroimaging.

We propose a generalized expectation-maximization algorithm (GEM; Dempster et al., 1977)
to search for penalized maximum likelihood estimators \citep*{Rido97, Ciup03,Chen08}
 of the hidden Ising model parameters.
The penalized likelihood prevents
  the
 unboundedness
   of the likelihood function, and the proposed GEM uses Monte Carlo averages via Gibbs sampler
 \citep*{Gema84,Robe94} to overcome the intractability of computing the normalizing constant in the underlying Ising model.
Then the LIS-based FDR procedures can be conducted by plugging in the estimates of the
 hidden Ising model
parameters.
In what follows, we use the term ``HMRF" to refer to the 3D hidden Ising model.

The article is organized as follows.
In Section 2, we introduce the HMRF model, i.e., the hidden Ising model, for 3D imaging data. We provide the GEM algorithm for the HMRF parameter estimation and the implementation of the HMRF-LIS-based data-driven procedures in Section 3.
In Section 4, we conduct extensive simulations to compare the LIS-based procedures with conventional FDR methods. In Section 5, we apply the PLIS procedure to the
$^{18}\mbox{F}$-Fluorodeoxyglucose PET (FDG-PET)
image data of the Alzheimer's Disease Neuroimaging Initiative (ADNI),
which finds more signals than conventional methods.

\section{A Hidden Markov Random Field Model}
\label{s:hmrf}
Let $S$ be a finite lattice of $N$ voxels in an image grid, usually in a 3D space. Let $\bmath{\Theta}= \{ \Theta_s\in\{0,1\}: s \in S \}$ denote the set of latent states on $S$, where $\Theta_s=1$
if the null hypothesis at voxel $s$ is false and $\Theta_s=0$ otherwise.
For simplicity,
we follow \citet{Sun09} to call hypothesis $s$ to be nonnull if $\Theta_s=1$ and null otherwise. We also call voxel $s$ to be a signal if $\Theta_s=1$ and noise otherwise.
Let $\bmath{\Theta}$ be generated from a two-parameter Ising model with the following  probability distribution
\begin{align}
\label{Ising model}
P_{\bmath{\varphi}}(\bmath{\theta})
&=\frac{1}{Z({\bmath{\varphi}})}\exp\{\bmath{{\bmath{\varphi}}}^T\bmath{H}(\bmath{\theta})\}\nonumber\\
&=\frac{1}{Z(\beta,h)}\exp
\left\{
\beta\sum_{\langle s, t \rangle}\theta_s \theta_t+h\sum_{s\in S}\theta_s
\right\},
\end{align}
where $Z({\bmath{\varphi}})$ is the normalizing constant, ${\bmath{\varphi}}=(\beta,h)^T$, $\bmath{H}(\bmath{\theta})=(\sum_{\langle s,t \rangle}\theta_s \theta_t, \sum_{s\in S}\theta_s)^T$, and $\langle s, t \rangle$
denotes all the unordered pairs in $S$
 such that for any $s$,  $t$ is among the six nearest neighbors of voxel $s$ in a 3D setting. This model possesses the Markov property:
\begin{align*}
P_{\bmath{\varphi}}(\theta_s|\bmath{\theta}_{S\setminus \{s\}})
&=P_{\bmath{\varphi}}(\theta_s|\bmath{\theta}_{\mathcal{N}(s)})\nonumber\\
&=\frac{  \exp\{  \theta_s  (  \beta  \sum_{t\in \mathcal{N}(s)}\theta_t+h   )   \}  }{  1+\exp\{\beta\sum_{t\in \mathcal{N}(s)}\theta_t+h\}},
\end{align*}
where $S\setminus \{s\}$ denotes the set $S$ after removing $s$, and $\mathcal{N}(s) \subset S$ is the nearest neighborhood of $s$ in $S$.
Some parameter interpretations of $\beta$ and $h$ are given in Web Appendix A.

We assume the observed
$z$-values
$\bmath{X}=\{ X_s: s \in S \}$ are independent given $\bmath{\Theta}=\bmath{\theta}$ with
\begin{equation}
\label{P(X|Theta)}
P_{\bmath{\phi}}(\bmath{x}|\bmath{\theta})=\prod_{s\in S} P_{\bmath{\phi}}(x_s|\theta_s),
\end{equation}
where $P_{\bmath{\phi}}(x_s|\theta_s)$ denotes the following distribution
\begin{equation}
\label{P(X_i|Theta_i)}
X_s|\Theta_s \sim (1-\Theta_s)N(\mu_0,\sigma_0^2)+\Theta_s\sum_{l=1}^Lp_lN(\mu_l,\sigma_l^2)
\end{equation}
with
$(\mu_0,\sigma_0^2)=(0,1) $,
unknown parameters
${\bmath{\phi}}=(\mu_1, \sigma_1^2, p_1,...,\mu_L, \sigma_L^2, p_L)^T$,
$\sum_{l=1}^Lp_l=1$
and $p_l\ge 0$.
In particular, the
$z$-value
 $X_s$ follows the standard normal distribution under the null,
and the nonnull distribution is set to be the normal mixture that can be used to approximate a large collection of distributions \citep{Magd96,Efro04}.  The number of components $L$ in the nonnull distribution may be selected by, for example, the Akaike or Bayesian information criterion. Following the recommendation of \citet{Sun09}, we use $L=2$ for the ADNI image analysis.

Markov random fields (MRFs; Bremaud, 1999) are a natural generalization of Markov chains (MCs), where the time index of MC is replaced by the space index of MRF. It is well known that any one-dimensional MC is an MRF, and any one-dimensional stationary finite-valued MRF is an MC \citep{Chan14}. When $S$ is taken to be one-dimensional, the above approach based on \eqref{Ising model}-\eqref{P(X_i|Theta_i)} reduces to the HMC method of \citet{Sun09}.

\section{Hidden Markov Random Field LIS-Based FDR Procedures}
\label{s:hmrf-fdr}
\citet{Sun09} developed a compound decision theoretic framework for multiple testing under HMC dependence and proposed LIS-based
oracle and data-driven
testing procedures that aim to minimize the FNR subject to a constraint on FDR. We extend these procedures under HMRF for image data. The oracle LIS for hypothesis $s$ is defined as $LIS_s(\bmath{x})=P_{\bmath{\Phi}}(\Theta_s=0|\bmath{x})$ for a given
parameter vector
${\bmath{\Phi}}$.
In our model, ${\bmath{\Phi}}=({\bmath{\phi}}^T,{\bmath{\varphi}}^T)^T$.
Let $LIS_{(1)}(\bmath{x}),...,LIS_{(N)}(\bmath{x})$ be the ordered LIS values and $\mathcal{H}_{(1)},...,\mathcal{H}_{(N)}$ the corresponding null hypotheses. The oracle procedure operates as follows: for a prespecified FDR level $\alpha$,
\begin{align}
\label{OR procedure}
&\text{let} \ k=\max \left\{i: \frac{1}{i}\sum_{j=1}^iLIS_{(j)}(\bmath{x})\le \alpha\right\},\nonumber\\
&\text{then reject all}\ \mathcal{H}_{(i)}, \, i=1,...,k.
\end{align}
Parameter ${\bmath{\Phi}}$ is unknown in practice. We can use the data-driven procedure that simply replaces $LIS_{(i)}(\bmath{x})$ in \eqref{OR procedure} with $\widehat{LIS}_{(i)}(\bmath{x})=P_{\hat{{\bmath{\Phi}}}}(\Theta_{(i)}=0|\bmath{x})$, where ${\hat{{\bmath{\Phi}}}}$ is an estimate of ${\bmath{\Phi}}$.

If all the tests are partitioned into multiple groups and each group follows its own HMRF,
in contrast to the separated LIS (SLIS) procedure that conducts the LIS-based FDR procedure separately for each group at the same FDR level $\alpha$ and then combines the testing results, we follow \citet{Wei09} to propose a pooled LIS (PLIS) procedure that is more efficient in reducing the global FNR. The PLIS follows the same procedure as \eqref{OR procedure}, but with $LIS_{(1)},...,LIS_{(N)}$ being the ordered test statistics from all groups.

Note that the model homogeneity, which is required in \citet{Sun09} and \citet{Wei09} for HMCs, fails to hold for the HMRF model. In other words,
$P(\Theta_s=1)$ for the interior voxels with six nearest neighbors are different to those for the boundary voxels with less than six nearest neighbors. We show the validity and optimality of the oracle HMRF-LIS-based procedures in Web Appendix B.

We now provide details of the LIS-based data-driven procedure for 3D image data, where the parameters of the HMRF model need to be estimated from observed test data.

\subsection{A Generalized EM Algorithm}

The observed likelihood function under HMRF, $L(\bmath{\Phi}|\bmath{x})=P_{\bmath{\Phi}}(\bmath{x})=
\sum_{\bmath{\Theta}}P_{\bmath{\phi}}(\bmath{x}|\bmath{\Theta})P_{\bmath{\varphi}}(\bmath{\Theta})$,
is unbounded (see Web Appendix C for details).
One solution to avoid the
unboundedness
is to replace the likelihood by a penalized likelihood \citep{Rido97,Ciup03}
\begin{equation}
\label{penalized likelihood}
pL({\bmath{\Phi}}|\bmath{x})=L({\bmath{\Phi}}|\bmath{x})\prod_{l=1}^Lg(\sigma_l^2),
\end{equation}
where $g(\sigma_l^2)$, $l=1,\dots,L$, are penalty functions that ensure the boundedness of $pL({\bmath{\Phi}}|\bmath{x})$. We follow \cite{Rido97} and \cite{Ciup03} to choose
\[
g(\sigma_l^2)\propto\frac{1}{\sigma_l^{2b}}\exp\left\{-\frac{a}{\sigma_l^2}\right\}, \quad a>0, b\ge 0,
\]
where $x\propto y$ means that $x=cy$ with a positive constant $c$ independent of any parameter.
Note that \eqref{penalized likelihood} reduces to the unpenalized likelihood function when $a=b=0$.
When $a>0$ and $b>1$, the penalized likelihood approach is equivalent to setting $g(\sigma_l^2)$ to be the inverse gamma distribution, which is a classical prior distribution for the variance of a normal distribution in Bayesian statistics \citep{Hoff09}.
We do not
impose any prior distribution here.
The choice of $a$ and $b$ does not impact the strong consistency of the penalized maximum likelihood estimator (PMLE) based on the same penalty function for a finite mixture of normal distributions \citep{Ciup03,Chen08}.
Such a penalty
performs well in the simulations, though
formal proof of the consistency of PMLE for hidden Ising model remains an open question.

We develop an EM algorithm based on the penalized likelihood \eqref{penalized likelihood} for the estimation of parameters in the HMRF model characterized by \eqref{Ising model}-\eqref{P(X_i|Theta_i)}.
We introduce unobservable
categorical variables $\bmath{K}=\left\{K_s:s\in S  \right\}$, where $K_s=0$ if $\Theta_s=0$, and $K_s\in\{1,...,L\}$ if $\Theta_s=1$.
Hence, $P(K_s{=}0|\Theta_s{=}0)=1$ and we denote $P(K_s{=}l|\Theta_s{=}1)=p_l$.
From \eqref{P(X_i|Theta_i)}, we let
$X_s|K_s \sim N(\mu_{K_s},\sigma^2_{K_s})$.
To estimate the HMRF parameters ${\bmath{\Phi}}=( {\bmath{\phi}}^T,{\bmath{\varphi}}^T)^T$, $(\bmath{\Theta},\bmath{K},\bmath{X})$ are used as the complete data variables
to construct the auxiliary function in the $(t+1)$st iteration of EM algorithm given the observed data $\bmath{x}$ and the current estimated parameters ${\bmath{\Phi}}^{(t)}$:
\begin{equation*}
Q({\bmath{\Phi}}|{\bmath{\Phi}}^{(t)})=E_{{\bmath{\Phi}}^{(t)}}[\log P_{\bmath{\Phi}}(\bmath{\Theta},\bmath{K},\bmath{X})|\bmath{x}]+\sum_{l=1}^L\log g(\sigma_l^2),
\end{equation*}
where $P_{\bmath{\Phi}}(\bmath{\Theta},\bmath{K},\bmath{X})=P_{\bmath{\varphi}}(\bmath{\Theta})P_{\bmath{\phi}}(\bmath{X},\bmath{K}|\bmath{\Theta})$ $=P_{\bmath{\varphi}}(\bmath{\Theta})\prod_{s\in S}P_{\bmath{\phi}}(X_s,K_s|\Theta_s)$.
The $Q$-function can be further written as follows
\[
Q({\bmath{\Phi}}|{\bmath{\Phi}}^{(t)})=Q_1({\bmath{\phi}}|{\bmath{\Phi}}^{(t)})+Q_2({\bmath{\varphi}}|{\bmath{\Phi}}^{(t)}),
\]
where
\begin{align*}
Q_1({\bmath{\phi}}|{\bmath{\Phi}}^{(t)})&=\sum_{\bmath{\Theta}}\sum_{\bmath{K}}P_{{\bmath{\Phi}}^{(t)}}(\bmath{\Theta},\bmath{K}|\bmath{x})\log P_{\bmath{\phi}}(\bmath{x},\bmath{K}|\bmath{\Theta})\nonumber\\
&\qquad+\sum_{l=1}^L\log g(\sigma_l^2)
\end{align*}
and
\begin{equation*}
Q_2({\bmath{\varphi}}|{\bmath{\Phi}}^{(t)})=\sum_{\bmath{\Theta}}P_{{\bmath{\Phi}}^{(t)}}(\bmath{\Theta}|\bmath{x})\log P_{\bmath{\varphi}}(\bmath{\Theta}).
\end{equation*}
Therefore, we can maximize $Q({\bmath{\Phi}}|{\bmath{\Phi}}^{(t)})$ for ${\bmath{\Phi}}$ by maximizing $Q_1({\bmath{\phi}}|{\bmath{\Phi}}^{(t)})$ for ${\bmath{\phi}}$ and $Q_2({\bmath{\varphi}}|{\bmath{\Phi}}^{(t)})$ for ${\bmath{\varphi}}$, separately.

Maximizing $Q_1({\bmath{\phi}}|{\bmath{\Phi}}^{(t)})$ under the constraint $\sum_{l=1}^Lp_l=1$ by the method of Lagrange multipliers yields,
\begin{eqnarray}
\label{pEll estimate}
p_{l}^{(t+1)}&=&\frac{\sum_{s \in S}w_s^{(t)}(l)}{\sum_{s \in S}\gamma_s^{(t)}(1)}, \\
\label{mu estimate}
\mu_l^{(t+1)}&=&\frac{\sum_{s \in S}w_s^{(t)}(l)x_s}{\sum_{s \in S}w_s^{(t)}(l)}, \\
\label{sigma estimate}
(\sigma_l^2)^{(t+1)}&=&\frac{2a+\sum_{s \in S}w_s^{(t)}(l)(x_s-\mu_l^{(t+1)})^2}{2b+\sum_{s \in S}w_s^{(t)}(l)},
\end{eqnarray}
where
\begin{eqnarray*}
w_s(l)&=&\frac{\gamma_s(1)p_{l}f_{l}(x_s)}{f(x_s)},
\\
\gamma_s(i)&=&P_{{\bmath{\Phi}}}(\Theta_s=i|\bmath{x}),
\\
f_l&=&N(\mu_l,\sigma_l^2), \quad \text{and} \  f=\sum_{l=1}^L p_lf_l.
\end{eqnarray*}
For $Q_2({\bmath{\varphi}}|{\bmath{\Phi}}^{(t)})$,  taking its first and second derivatives with respect to ${\bmath{\varphi}}$, we obtain
\begin{eqnarray*}
\label{U function}
\bmath{U}^{(t+1)}({\bmath{\varphi}})
&=& \frac{\partial}{\partial {\bmath{\varphi}}}Q_2({\bmath{\varphi}}|{\bmath{\Phi}}^{(t)})\nonumber\\
&=&E_{{\bmath{\Phi}}^{(t)}}[\bmath{H}(\bmath{\Theta})|\bmath{x}]-E_{\bmath{\varphi}}[\bmath{H}(\bmath{\Theta})],
\\
\label{I function}
\bmath{I}({\bmath{\varphi}}) &=& - \frac{\partial^2}{\partial {\bmath{\varphi}} \partial {\bmath{\varphi}}^T}Q_2({\bmath{\varphi}}|{\bmath{\Phi}}^{(t)})=Var_{\bmath{\varphi}}[\bmath{H}(\bmath{\Theta})].
\end{eqnarray*}
Maximizing $Q_2({\bmath{\varphi}}|{\bmath{\Phi}}^{(t)})$ is then equivalent to solving the nonlinear equation:
\begin{equation}
\label{Q2 nonlinear equation}
\bmath{U}^{(t+1)}({\bmath{\varphi}})=E_{{\bmath{\Phi}}^{(t)}}[\bmath{H}(\bmath{\Theta})|\bmath{x}]-E_{\bmath{\varphi}}[\bmath{H}(\bmath{\Theta})]=\bmath{0}.
\end{equation}

It can be shown that equation \eqref{Q2 nonlinear equation} has a unique solution and can be solved by the Newton-Raphson (NR) method \citep{Stoe02}. However, a starting point that is not close enough to the solution may result in divergence of  the NR method. Therefore, rather than searching for the solution of equation \eqref{Q2 nonlinear equation} over all ${\bmath{\varphi}}$, we choose a ${\bmath{\varphi}}^{(t+1)}$ that increases $Q_2({\bmath{\varphi}}|{\bmath{\Phi}}^{(t)})$ over its value at ${\bmath{\varphi}}={\bmath{\varphi}}^{(t)}$. Together with the maximization of $Q_1({\bmath{\phi}}|{\bmath{\Phi}}^{(t)})$, the approach leads to $Q({\bmath{\Phi}}^{(t+1)}|{\bmath{\Phi}}^{(t)}) \ge Q({\bmath{\Phi}}^{(t)}|{\bmath{\Phi}}^{(t)})$ and thus $pL({\bmath{\Phi}}^{(t+1)}|\bmath{x})\ge pL({\bmath{\Phi}}^{(t)}|\bmath{x})$, which is termed a GEM algorithm \citep{Demp77}. To find such a ${\bmath{\varphi}}^{(t+1)}$ that increases the $Q_2$-function, a backtracking line search algorithm \citep{Noce06} is applied with a set of decreasing positive values $\lambda_m$ in the following
\begin{equation}
\label{backtrack}
{\bmath{\varphi}}^{(t+1,m)}={\bmath{\varphi}}^{(t)}+\lambda_m\bmath{I}({\bmath{\varphi}}^{(t)})^{-1}\bmath{U}^{(t+1)}({\bmath{\varphi}}^{(t)}),
\end{equation}
where $m=0,1, ...,$ and ${\bmath{\varphi}}^{(t+1)}={\bmath{\varphi}}^{(t+1,m)}$ which is the first one satisfying the Armijo condition \citep{Noce06}
\begin{align}
\label{Armijo condition}
\lefteqn{Q_2({\bmath{\varphi}}^{(t+1,m)}|{\bmath{\Phi}}^{(t)})-Q_2({\bmath{\varphi}}^{(t)}|{\bmath{\Phi}}^{(t)})}\nonumber\\
&\qquad \ge \alpha \lambda_m\bmath{U}^{(t+1)}({\bmath{\varphi}}^{(t)})^T\bmath{I}({\bmath{\varphi}}^{(t)})^{-1}\bmath{U}^{(t+1)}({\bmath{\varphi}}^{(t)}).
\end{align}
Since $\bmath{I}({\bmath{\varphi}}^{(t)})$ is positive-definite, the Armijo condition guarantees the increase of $Q_2$-function. In practice, $\alpha$ is chosen to be quite small. We adopt $\alpha=10^{-4}$, which is recommended by \citet{Noce06}, and halve the Newton-Raphson step length each time by using $\lambda_m=2^{-m}$.

In the GEM algorithm, Monte Carlo averages are used via Gibbs sampler to approximate the quantities of interest that are involved with the intractable normalizing constant of the Ising model.
By the ergodic theorem of the Gibbs sampler \citep{Robe94} (see Web Appendix D for details),
\begin{eqnarray*}
\bmath{U}^{(t+1)}({\bmath{\varphi}})&\approx&\frac{1}{n}\sum_{i=1}^n
\left(
\bmath{H}(\bmath{\theta}^{(t,i,\bmath{x})})-\bmath{H}(\bmath{\theta}^{(i,{\bmath{\varphi}})})
\right),
\\
\bmath{I}({\bmath{\varphi}})&\approx&
\frac{1}{n-1}\sum_{i=1}^n \left(
\bmath{H}(\bmath{\theta}^{(i, {\bmath{\varphi}})})-\frac{1}{n} \sum_{j=1}^n \bmath{H}(\bmath{\theta}^{(j, {\bmath{\varphi}})})
\right)^{\bigotimes 2},
\end{eqnarray*}
where $\{\bmath{\theta}^{(t, 1, \bmath{x})}, ..., \bmath{\theta}^{(t, n, \bmath{x})}\}$
are large $n$ samples
successively
generated by the Gibbs sampler from
\[
P_{{\bmath{\Phi}}^{(t)}}(\bmath{\theta}|\bmath{x})=\frac{\exp
\left\{
\beta^{(t)}\sum_{\langle s, r \rangle}\theta_s \theta_r+\sum_{s\in S}h_s^{(t)}\theta_s
\right\}}{Z\left(\beta^{(t)},\{h_s^{(t)}\}_{s\in S}\right)},
\]
with
\begin{align*}
h_s^{(t)}&=
h^{(t)}
-\log\left(
\frac{1}{\sqrt{2\pi\sigma_0^2}}\exp\left\{-\frac{(x_s-\mu_0)^2}{2\sigma_0^2}\right\}
\right)
\nonumber\\
& \qquad +
\log\left(
\sum_{l=1}^L\frac{p_l^{(t)}}{\sqrt{2\pi\sigma_l^{2^{(t)}}}}\exp\left\{
-\frac{(x_s-\mu_l^{(t)})^2}{2\sigma_l^{2^{(t)}}}
\right\}
\right)
\end{align*}
and $Z\left(\beta^{(t)},\{h_s^{(t)}\}_{s\in S}\right)$ being the normalizing constant,
and $\{\bmath{\theta}^{(1, {\bmath{\varphi}})}, ..., \bmath{\theta}^{(n, {\bmath{\varphi}})}\}$ are generated from $P_{\bmath{\varphi}}(\bmath{\theta})$. Here for vector $v$, $v ^{\otimes 2}=vv^T$.  Similarly,
\[
\frac{C}{Z({\bmath{\varphi}})}=E_{\bmath{\varphi}}[\exp\{-{\bmath{\varphi}}^T\bmath{H}(\bmath{\Theta})\}]\approx\frac{1}{n}\sum_{i=1}^n\exp\{-{\bmath{\varphi}}^T\bmath{H}(\bmath{\theta}^{(i,{\bmath{\varphi}})})\},
\]
where $C$ is the number of all possible configurations $\bmath{\theta}$ of $\bmath{\Theta}$. Then
the difference between $Q_2$-functions in the Armijo condition can be approximated by
\begin{align*}
\lefteqn{Q_2({\bmath{\varphi}}^{(t+1,m)}|{\bmath{\Phi}}^{(t)})-Q_2({\bmath{\varphi}}^{(t)}|{\bmath{\Phi}}^{(t)})}\\
&\qquad \approx
\frac{1}{n}({\bmath{\varphi}}^{(t+1,m)}-{\bmath{\varphi}}^{(t)})^T\sum_{i=1}^n\bmath{H}(\bmath{\theta}^{(t,i,\bmath{x})})\\
& \qquad \qquad +
\log \left(
\frac{\sum_{i=1}^n\exp\{-{{\bmath{\varphi}}^{(t+1,m)}}^T\bmath{H}(\bmath{\theta}^{(i,{\bmath{\varphi}}^{(t+1,m)})})\}}
{\sum_{i=1}^n\exp\{-{{\bmath{\varphi}}^{(t)}}^T\bmath{H}(\bmath{\theta}^{(i,{\bmath{\varphi}}^{(t)})})\}}
\right).
\end{align*}
Back to $Q_1({\bmath{\phi}}|{\bmath{\Phi}}^{(t)})$, the local conditional probability of $\bmath{\Theta}$ given $\bmath{x}$ can also be approximated by the Gibbs sampler:
\begin{equation}
\label{gamma function}
\gamma_s^{(t)}(i)=P_{{\bmath{\Phi}}^{(t)}}(\Theta_s=i|\bmath{x})\approx \frac{1}{n}\sum_{k=1}^n\bmath{1}(\theta_s^{(t,k,\bmath{x})}=i).
\end{equation}

\subsection{Implementation of the LIS-Based FDR Procedure}

The algorithm for the LIS-based data-driven procedure, denoted as LIS for single group analysis, SLIS for separate analysis of multiple groups, and PLIS for pooled analysis for multiple groups, is given below:

\medskip

1.  Set initial values ${\bmath{\Phi}}^{(0)}=\{{\bmath{\phi}}^{(0)},{\bmath{\varphi}}^{(0)}\}$ for the model parameters ${\bmath{\Phi}}$ of each group;

2.  Update ${\bmath{\phi}}^{(t)}$ from equations \eqref{pEll estimate}, \eqref{mu estimate} and
\eqref{sigma estimate};

3. Update ${\bmath{\varphi}}^{(t)}$ from equations \eqref{backtrack} and \eqref{Armijo condition};

4. Iterate Steps 2 and 3 until convergence, then obtain the estimate $\hat{{\bmath{\Phi}}}$ of ${\bmath{\Phi}}$;

5. Plug-in $\hat{{\bmath{\Phi}}}$ to obtain the test statistics $\widehat{LIS}$ from equation \eqref{gamma function};

6. Apply the data-driven procedure (LIS, SLIS or PLIS).

\medskip

The GEM algorithm is stopped when the following stopping rule
\begin{equation}
\label{stopping rule}
\max_i\left( \frac{|\Phi_i^{(t+1)}-\Phi_i^{(t)}|}{|\Phi_i^{(t)}|+\epsilon_1}        \right)<\epsilon_2,
 \end{equation}
where $\Phi_i$ is the $i$th coordinate of vector ${\bmath{\Phi}}$, is satisfied for three consecutive regular Newton-Raphson iterations with $m=0$ in \eqref{backtrack}, or the prespecified maximum number of iterations is reached.
Stopping rule \eqref{stopping rule} was applied by \citet{Boot99} to the Monte Carlo EM method, where they set $\epsilon_1=0.001$, $\epsilon_2$ between 0.002 and 0.005, and the rule to be satisfied for three consecutive iterations to avoid stopping the algorithm prematurely because of Monte Carlo error. We used $\epsilon_1=\epsilon_2=0.001$ in simulation studies and real-data analysis. Constant $\alpha=10^{-4}$ is recommended by \citet{Noce06} for the Armijo condition \eqref{Armijo condition}, and the Newton-Raphson step length in \eqref{backtrack} is halved by using $\lambda_m=2^{-m}$ . In practice, the Armijo condition \eqref{Armijo condition} might not be satisfied when the step length $\| {\bmath{\varphi}}^{(t+1,m)}-{\bmath{\varphi}}^{(t)}\|$ is very small. In this situation, the iteration within Step 3 is stopped by an alternative criterion
 \[
\max_i\left( \frac{|\varphi_i^{(t+1,m)}-\varphi_i^{(t)}|}{|\varphi_i^{(t)}|+\epsilon_1}        \right)<\epsilon_3
 \]
 with $\epsilon_3<\epsilon_2$, for example, $\epsilon_3=10^{-4}$ if $\epsilon_2=0.001$.
Small $a$ and $b$ should be chosen in \eqref{sigma estimate}. We choose $a=1$ and $b=2$.

\section{Simulation Studies}
\label{s:simulations}
The simulation setups are similar to those in \citet{Sun09} and \citet{Wei09}, but with 3D data.
The performances of the proposed LIS-based oracle (OR) and data-driven procedures are compared with the BH approach \citep{Benj95},
the $q$-value procedure \citep{Store03},
and the local FDR (Lfdr) procedure \citep{Sun07} for single group analysis; and the performances of SLIS and PLIS are compared with BH, $q$-value, and the conditional Lfdr (CLfdr) procedure \citep{Cai09} for multiple groups. The Lfdr and CLfdr procedures are shown to be optimal for independent tests \citep{Sun07,Cai09}.
For simulations with multiple groups, all the procedures are globally implemented using all the locally computed test statistics based on each method from each group. The $q$-values are obtained using the
R package {\tt qvalue} \citep{Dabn14}.
For the Lfdr or CLfdr procedure, we use the proportion of the null cases generated from the Ising model with given parameters as the estimate of the probability of the null cases $P(\Theta_s=0)$, together with the given null and nonnull distributions without estimating their parameters. For the LIS-based data-driven procedures, the maximum number of GEM iterations is set to be 1,000 with
$\epsilon_1=\epsilon_2=0.001$, $\epsilon_3=\alpha=10^{-4}$, $a=1$ and $b=2$.
For the Gibbs sampler, 5,000 samples are generated from 5,000 iterations after a burn-in period of 1,000 iterations. In all simulations,
each HMRF is
on a $N=15{\times}15{\times}15$ cubic lattice $S$, the number of replications $M=200$ is the same as that in \citet{Wei09}, and the nominal FDR level is set at 0.10.

\subsection{Single-Group Analysis}
\subsubsection{Study 1: $L=1$}
The MRF $\bmath{\Theta}=\{\Theta_s:s\in S\}$ is generated from the Ising model \eqref{Ising model} with parameters $(\beta,h)$, and the observations $\bmath{X}=\{X_s:s\in S\}$ are generated conditionally on $\bmath{\Theta}$ from $X_s|\Theta_s \sim (1-\Theta_s)N(0,1)+\Theta_sN(\mu_1,\sigma_1^2)$.
Note that the MRF $\bmath{\Theta}$ is not observable in practice.
Figure 1 shows the comparisons of the performance of  BH,
$q$-value,
Lfdr, OR and LIS. In Figure 1(1a-1c), we fix $h=-2.5$, set $\mu_1=2$ and $\sigma_1^2=1$, and plot FDR, FNR, and the average number of true positives (ATP) yielded by these procedures as functions of $\beta$. In Figure 1(2a-2c), we fix $\beta=0.8$, set $\mu_1=2$ and $\sigma_1^2=1$, and plot FDR, FNR and ATP as functions of $h$. In Figure 1(3a-3c), we fix  $\beta=0.8$ and $h=-2.5$, set $\sigma_1^2=1$, and plot FDR, FNR and ATP as functions of $\mu_1$.
The corresponding average proportions of the nulls, denoted by $P_0$, for each Ising model are given in Figure 1(1d-3d).
The initial values for the numerical algorithm are set at $\beta^{(0)}=h^{(0)}=0,\mu_1^{(0)}=\mu_1+1$ and $\sigma_1^{2{(0)}}=2$.

\begin{figure}
\label{fig1}
\centering
\includegraphics[width=16cm,height=12cm]{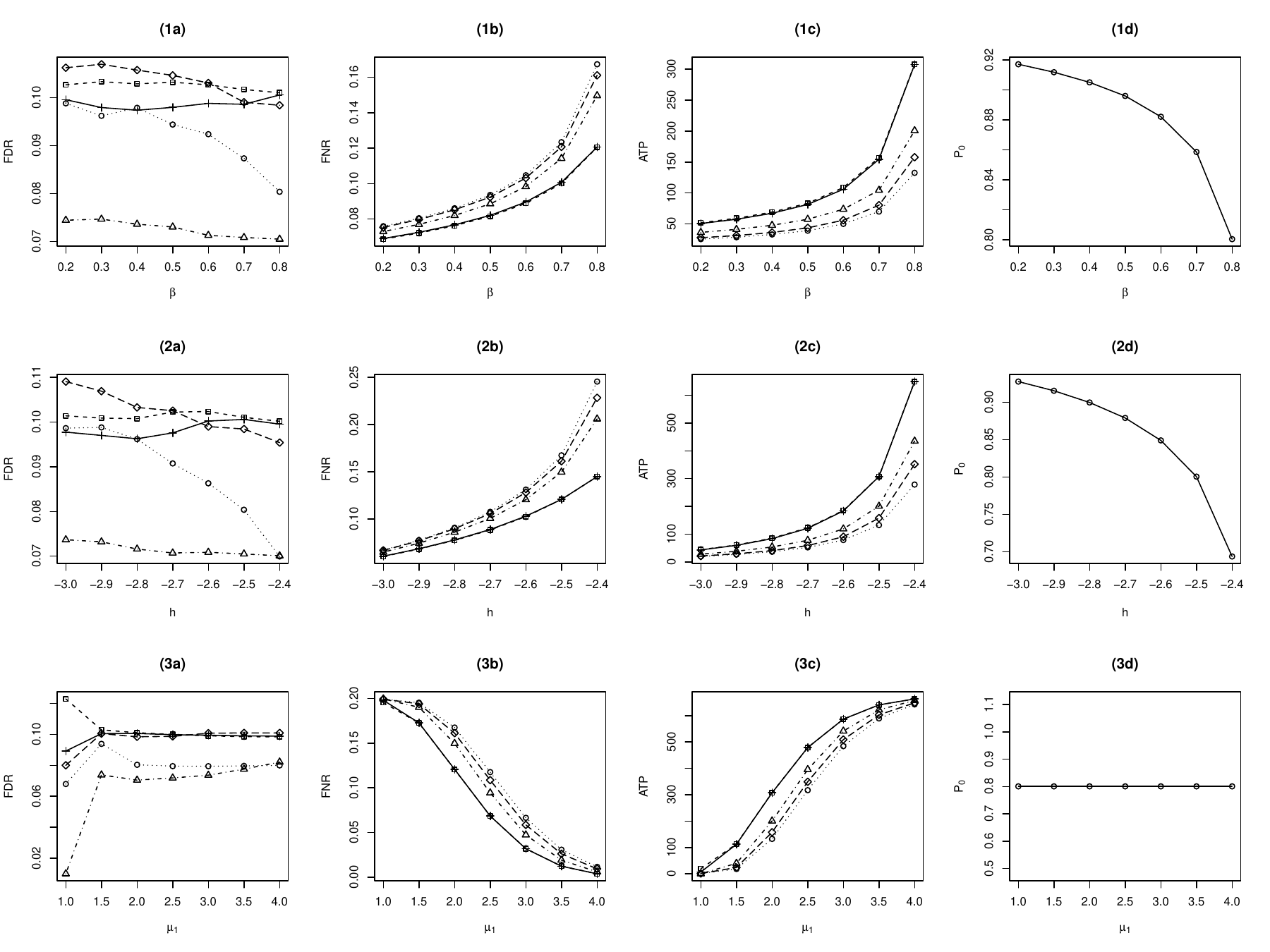}
\caption{Comparison of BH ($\bigcirc$), $q$-value ($\Diamond$), Lfdr ($\triangle$), OR ($+$) and LIS ($\square$) for a single group with $L=1$.}
\end{figure}

From Figure 1(1a-3a), we can see that the FDR levels of all
five
procedures are controlled around 0.10 except one case of the LIS procedure in Figure 1(3a) with the lowest $\mu_1$, whereas the BH and Lfdr procedures are generally conservative. This case of obvious deviation of the LIS procedure is likely caused by the small lattice size $N$. As a confirmation, additional simulations by increasing the lattice size $N$ to $30 {\times} 30 {\times} 30$ yield an FDR of 0.1019 for the same setup. From Figure 1(1b-3b) and (1c-3c) we can see that the two curves of OR and LIS procedures are almost identical, indicating that the data-driven LIS procedure works equally well as the OR procedure. These plots also show that the LIS procedure outperforms
BH, $q$-value and Lfdr procedures
with increased margin of performance in FNR and ATP as $\beta$ or $h$ increases or $\mu_1$ is at a moderate level. Note that from
Web Appendix A, we can see that $\beta$ controls how likely the same-state cases cluster together, and $(\beta,h)$ together control the proportion of the aggregation of nonnulls relative to that of nulls.

\subsubsection{Study 2: $L=2$}
We now consider the case where the nonnull distribution is a mixture of two normal distributions. The MRF is generated from the Ising model \eqref{Ising model} with fixed parameters $\beta=0.8$ and $h=-2.5$,
and the nonnull distribution is a two-component normal mixture
$p_1N(\mu_1,\sigma_1^2)+p_2N(\mu_2,\sigma_2^2)$ with  fixed $p_1=p_2=0.5$, $\mu_2=2,$ and $\sigma_2^2=1$. In Figure 2(1a-1c), $\sigma_1^2$ varies from 0.125 to 8, and $\mu_1=-2$. In Figure 2(2a-2c), we fix $\sigma_1^2=1$ and
vary
$\mu_1$ from $-4$ to $-1$. The initial values are set at $\beta^{(0)}=h^{(0)}=0$, $p_1^{(0)}=1-p_2^{(0)}=0.3$, $\mu_l^{(0)}=\mu_l+1,$ and $\sigma_l^{2{(0)}}=\sigma_l^2+1, l=1,2$.

\begin{figure}
\label{fig2}
\centering
\includegraphics[width=15cm,height=15cm]{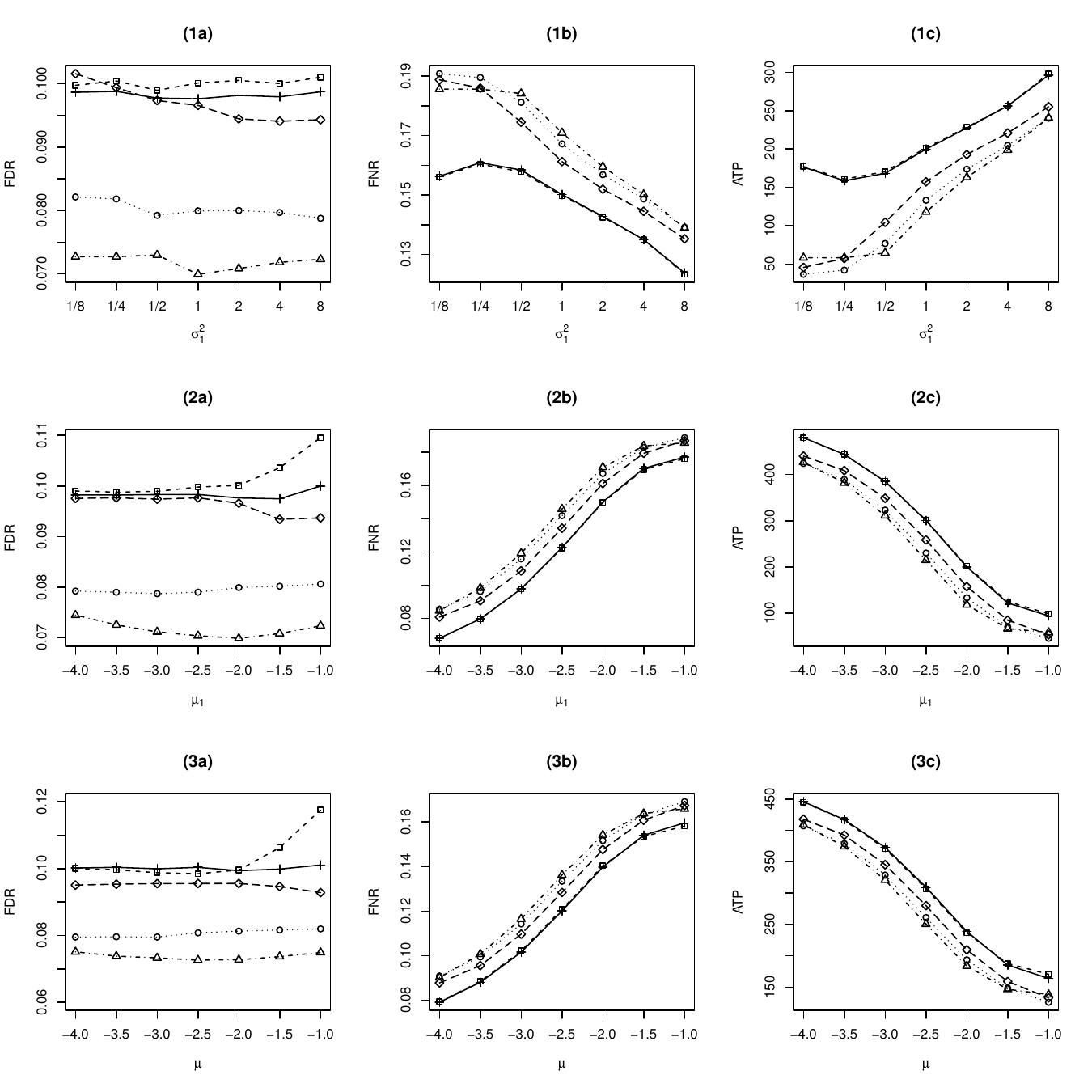}
\caption{Comparison of BH ($\bigcirc$), $q$-value ($\Diamond$), Lfdr ($\triangle$), OR ($+$) and LIS ($\square$) for a single group with $L=2$ (see 1a-2c), and the one with $L$ being misspecified (see 3a-3c).}
\end{figure}

Similar to Figure 1, we can see that the FDR levels of
all the procedures are controlled around 0.10, where BH and Lfdr are conservative, and OR and LIS perform similarly and outperform
the other three procedures.
In Figure 2(2a) at $\mu_{1}=-1$,
additional simulations yield an FDR of 0.1035 when the lattice size $N$ is increased to $30 {\times} 30 {\times} 30$ for the same setup.

The results from both simulation studies are very similar to
those
in \citet{Sun09} for the one-dimensional case using HMC. It is clearly seen that, for dependent tests, incorporating dependence structure into a multiple-testing procedure improves efficiency dramatically.

\vspace{-0.15in}

\subsubsection{Study 3: misspecified nonnull}
Following \citet{Sun09}, we consider the true
nonnull distribution to be the three-component normal mixture $0.4N(\mu,1)+0.3N(1,1)+0.3N(3,1)$, but use a misspecified two component normal mixture $p_1N(\mu_1,\sigma_1^2)+p_2N(\mu_2,\sigma_2^2)$ in the LIS procedure. The unobservable states are generated from the Ising model \eqref{Ising model} with fixed parameters $\beta=0.8$ and $h=-2.5$. The simulation results are displayed in Figure 2(3a-3c), the true $\mu$ varies from $-4$ to $-1$ with increments of size 0.5. The initial values are set at $\beta^{(0)}=h^{(0)}=0$, $p_1^{(0)}=p_2^{(0)}=0.5$, $\mu_1^{(0)}=-\mu_2^{(0)}=-2,$ and $\sigma_l^{2{(0)}}=2, l=1,2$.

Figure 2(3a-3c) shows that the LIS procedure performs similarly to OR under misspecified model. Additionally, the obvious biased FDR level by the LIS procedure at $\mu=-1$ reduces to 0.1067 when the lattice size $N$ is increased to $30 {\times} 30 {\times} 30$.

\vspace{-0.15in}

\subsection{Multiple-Group Analysis}
Voxels in a human brain can be naturally grouped into multiple functional regions. For simulations with grouped multiple tests, we consider two lattice groups each with size $15{\times} 15{\times} 15$.
The corresponding MRFs $\bmath{\Theta}_1=\{\Theta_{1s}:s\in S\}$ and $\bmath{\Theta}_2=\{\Theta_{2s}:s\in S\}$ are generated from the Ising model \eqref{Ising model} with parameters $(\beta_1=0.2,h_1=-1)$ and $(\beta_2=0.8,h_2=-2.5)$, respectively. The observations $\bmath{X}_k=\{X_{ks},s\in S\}$ are generated conditionally on $\bmath{\Theta}_k$, $k=1,2$, from $X_{ks}|\Theta_{ks} \sim (1-\Theta_{ks})N(0,1)+\Theta_{ks}N(\mu_k,\sigma_k^2)$, where $\mu_1$ varies from 1 to 4 with increments of size 0.5, $\mu_2=\mu_1+1$ and $\sigma_1^2=\sigma_2^2=1$. The initial values are
$\beta_1^{(0)}=\beta_2^{(0)}=h_1^{(0)}=h_2^{(0)}=0,$ $\mu_2^{(0)}=\mu_1^{(0)}=\mu_1+1,$ and $\sigma_1^{2^{(0)}}=\sigma_2^{2^{(0)}}=2$.

The simulation results are presented in Figure 3, which are similar to that in \citet{Wei09} for the one-dimensional case with multiple groups using HMCs. Figure 3(a) shows that all procedures are valid in controlling FDR at the prespecified level of 0.10, whereas BH and CLfdr procedures are conservative.
We also plot the within-group FDR levels of PLIS for each group separately. One can see that in order to minimize the global FNR level, the PLIS procedure may automatically adjust the FDRs of each individual group, either inflated or deflated reflecting the group heterogeneity, while the global FDR is appropriately controlled. In Figure 3(b) and (c) we
 can
see that both SLIS and PLIS outperform
 BH, $q$-value and CLfdr procedures,
indicating that utilizing the dependency information can improve the efficiency of a testing procedure, and the improvement is more evident for weaker signals (smaller values of $\mu_1$). Between the two LIS-based procedures, PLIS slightly outperforms SLIS, indicating the benefit of ranking the LIS test statistics globally.
In particular, ATP is 8.3\% higher for PLIS than for SLIS when $\mu_1=1$.
\begin{figure}
\label{fig3}
\centering
\includegraphics[width=15cm,height=5cm]{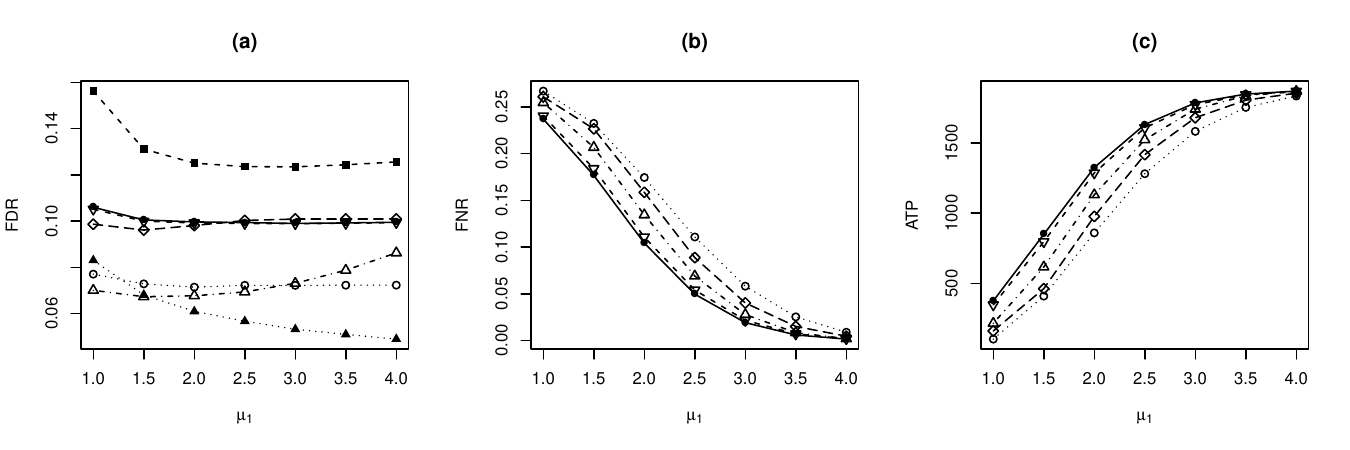}
\caption{Comparison of BH ($\bigcirc$), $q$-value ($\Diamond$), CLfdr ($\triangle$), SLIS ($\bigtriangledown$) and PLIS ($\bullet$) for two groups with $L=1$. In (a), $\blacksquare$ and $\blacktriangle$ represent the results by PLIS for each individual group; for PLIS, while the global FDR is controlled, individual-group FDRs may vary.}
\end{figure}

\vspace{-0.15in}
\section{ADNI FDG-PET Image Data Analysis}
\label{real data}
Alzheimer's disease (AD) is the most common cause of dementia in the elderly population. Much progress has been made in the diagnosis of AD including clinical assessment and neuroimaging techniques. One such extensively used neuroimaging technique is FDG-PET imaging, which is used to evaluate the cerebral metabolic rate of glucose (CMRgl).
We consider the FDG-PET image data from the
ADNI
database ({\tt adni.loni.usc.edu}) as an illustrative example.

The data set consists of the baseline FDG-PET images of
102 normal control (NC) subjects and 206 patients with mild cognitive impairment (MCI), a prodromal stage of AD. Sixty one brain regions of interest (ROIs) are considered (see Web Appendix E for details), where the number of voxels in each region ranges from 149 to 20,680 with a median of 2,517. The total number of voxels of these 61
ROIs
is $N=251,500$. The goal is to identify voxels with reduced CMRgl in MCI patients comparing to NC.

We apply the HMRF-PLIS procedure to the ADNI data, and compare to BH,
$q$-value
and CLfdr procedures.
We implement the BH procedure globally for the  61 ROIs, whereas we treat each region as a group for the
$q$-value,
CLfdr and PLIS procedures. For the BH and $q$-value procedures, a total number of $N$ two-sample Welch's $t$-tests \citep{Welc47} are performed, and their corresponding two-sided $p$-values are obtained.  For the PLIS and CLfdr procedures, $z$-values are used as the observed data $\bmath{x}$, which are obtained from those $t$ statistics by the transformation $z_i={\Phi}^{-1}[G_0(t_i)]$, where ${\Phi}$ and $G_0$ are the cumulative distribution functions of the standard normal and the $t$ statistic, respectively.
The null distribution is assumed to be the standard normal distribution. The nonnull distribution is assumed to be a two-component normal mixture for PLIS.
The LIS statistics in the PLIS procedure are approximated by  $10^6$ Gibbs-sampler samples, and the Lfdr statistics in the CLfdr procedure are computed by using the R code of \citet{Sun07}.
All the four testing procedures are controlled at a
 nominal
FDR level of 0.001.
In the GEM algorithm for HMRF estimation,
the initial values for $\beta$ and $h$ in the Ising model are set to be zero. The initial values for the nonnull distributions are estimated from the signals claimed by BH at an FDR level of 0.1. The maximum number of GEM iterations is set to be 5,000 with $\epsilon_1=\epsilon_2=0.001$, $\epsilon_3=\alpha=10^{-4}$, $a=1$ and $b=2$. For the Gibbs sampler embedded in the GEM, 5,000 samples are generated from 5,000 iterations after a burn-in period of 1,000 iterations. In this data analysis, the GEM algorithm reaches the maximum iteration and is then claimed to be converged for five ROIs.
Among all 61 ROIs, the estimates of $\beta$ have a median of $1.57$ with the interquartile range of 0.36, and the estimates of $h$ have a median of $-3.71$ with the interquartile range of 1.52. Such magnitude of parameter variation supports the multi-region analysis of the ADNI FDG-PET image data because even a 0.1 difference in $\beta$ or $h$ can result in quite different Ising models, see Figure 1(1d) and (2d).

Figure 4 shows the $z$-values (obtained by comparing CMRgl values between NC and MCI) of all the signals claimed by each procedure.
Figure 5 summarizes the number of voxels that are claimed as signals by each procedure.
We can see that PLIS finds the largest number of signals and covers 91.5\%, 97.2\% and 99.9\% of signals detected by CLfdr, $q$-value and BH, respectively.
It is interesting to see that the PLIS procedure finds more than 17 times signals as BH, twice as many signals as $q$-value, and about 20\% more signals than the CLfdr procedure.

Detailed interpretations of the scientific findings are provided in Web Appendix E.

\begin{figure}
\centering
\includegraphics[width=15cm]{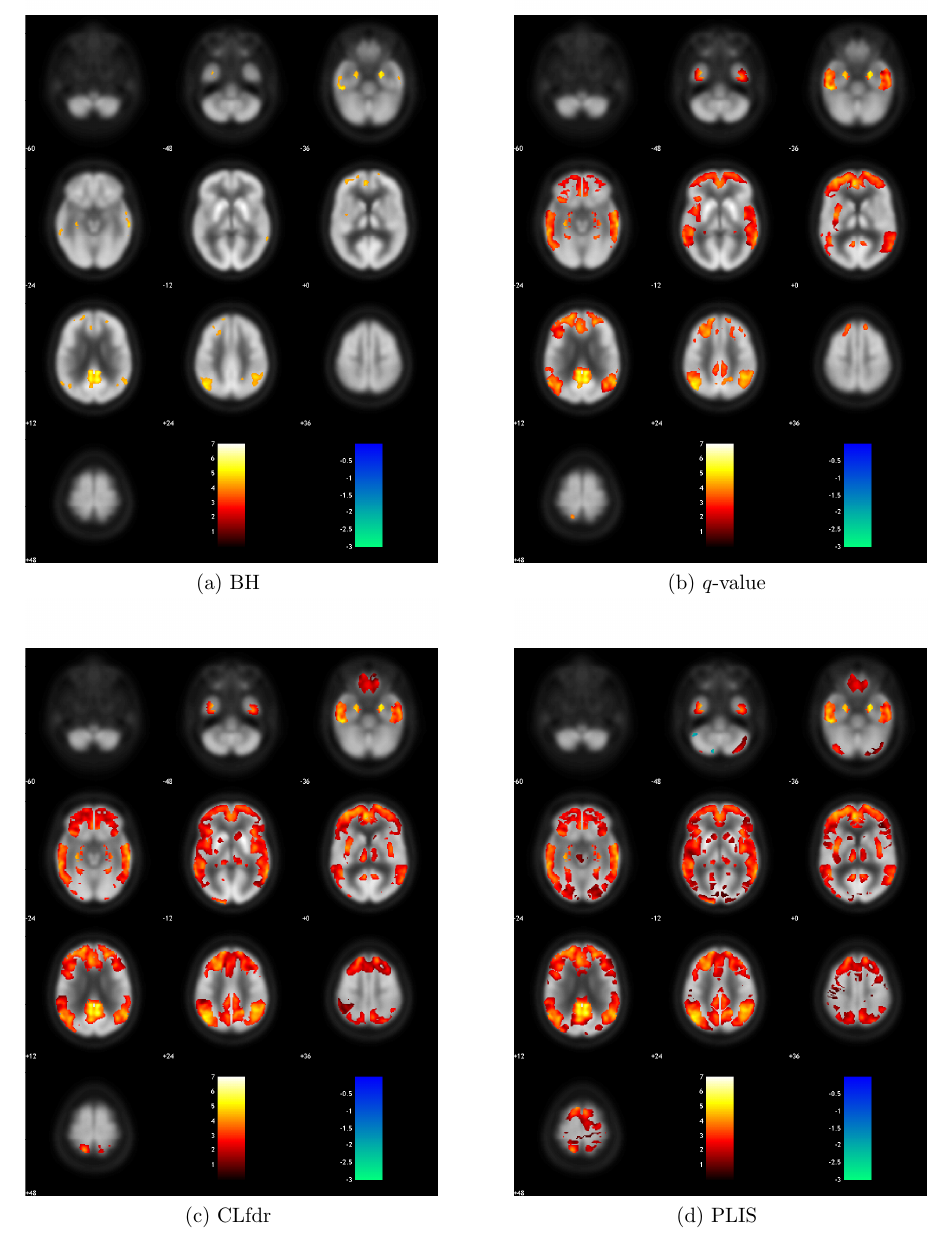}
\caption{$Z$-values of the signals found by each procedure for the comparison between NC and MCI.}
\end{figure}

\begin{figure}
\centering
\includegraphics[width=15cm]{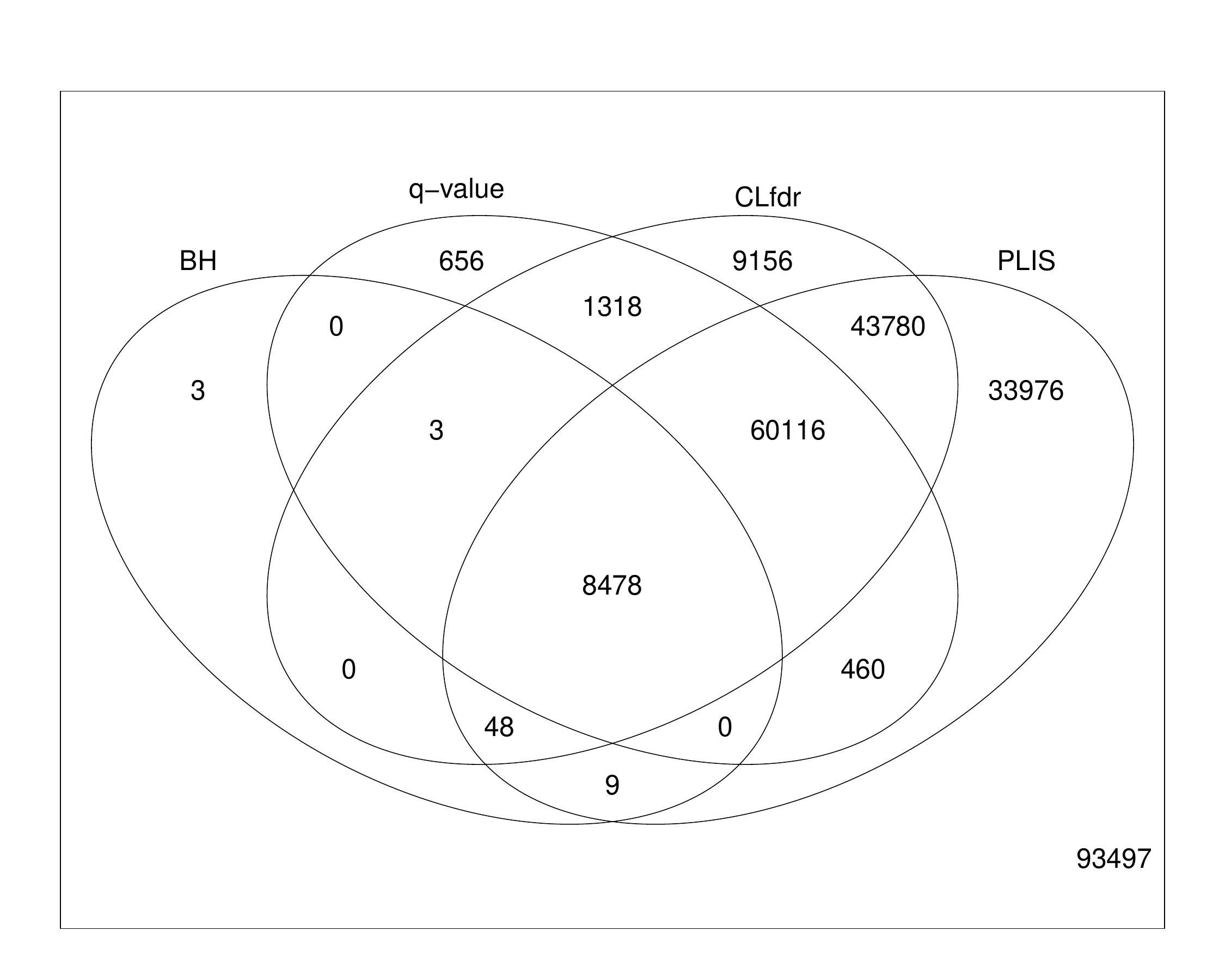}
\caption{Venn diagram for the number of signals found by each procedure for the comparison between NC and MCI.
Number of signals discovered by each procedure: BH=8,541, $q$-value=71,031, CLfdr=122,899, and PLIS=146,867.}
\end{figure}

\vspace{-0.2in}
\section{Concluding Remarks}
In this article, we consider LIS-based FDR procedures based on HMRF for 3D neuroimage data, where HMRF provides a natural way of modeling spatial correlations. The procedures aim to minimize the FNR while FDR is controlled at a prespecified level.  We find brain regions are spatially heterogeneous, hence model each region separately by a single HMRF,  and implement the PLIS procedure to minimize the global FNR. We propose a GEM algorithm based on the penalized likelihood
to obtain the
HMRF parameter estimates, which overcomes the unboundedness of the original likelihood function. Numerical analysis shows the superiority of the HMRF-LIS-based procedures over commonly used FDR procedures, illustrating the value of HMRF-LIS-based FDR procedures for spatially correlated image data. The asymptotic properties of the PMLE of HMRF and the data-driven HMRF-LIS-based procedures are of interest for future research.

\section{Supplementary Materials}
Web Appendix A mentioned in Sections~\ref{s:hmrf} and~\ref{s:simulations},
Web Appendices B-D referenced in Section~\ref{s:hmrf-fdr}, Web Appendix E mentioned in Section~\ref{real data},
and a MATLAB package implementing the proposed FDR procedure
are available with
this paper at the Biometrics website on Wiley Online
Library.

\backmatter
\section*{Acknowledgements}
We are grateful to Dr. Jeanine Houwing-Duistermaat, an Associate Editor and two anonymous referees for their helpful comments.
The research is supported in part by NIH grant R01-AG036802 and NSF grants DMS-1007590 and DMS-1407142.

We also would like to thank ADNI for providing the brain image data that were obtained from the ADNI database ({\tt adni.loni.usc.edu}). Data collection and sharing was funded by NIH grant U01-AG024904 and DOD grant W81XWH-12-2-0012. ADNI is funded by the National Institute on Aging, the National Institute of Biomedical Imaging and Bioengineering, and through generous contributions from the following: Alzheimer's Association; Alzheimer's Drug Discovery Foundation; Araclon Biotech; BioClinica, Inc.; Biogen Idec Inc.; Bristol-Myers Squibb Company; Eisai Inc.; Elan Pharmaceuticals, Inc.; Eli Lilly and Company; EuroImmun; F. Hoffmann-La Roche Ltd and its affiliated company Genentech, Inc.; Fujirebio; GE Healthcare; ; IXICO Ltd.; Janssen Alzheimer Immunotherapy Research \& Development, LLC.; Johnson \& Johnson Pharmaceutical Research \& Development LLC.; Medpace, Inc.; Merck \& Co., Inc.; Meso Scale Diagnostics, LLC.; NeuroRx Research; Neurotrack Technologies; Novartis Pharmaceuticals Corporation; Pfizer Inc.; Piramal Imaging; Servier; Synarc Inc.; and Takeda Pharmaceutical Company. The Canadian Institutes of Health Research is providing funds to support ADNI clinical sites in Canada. Private sector contributions are facilitated by the Foundation for the National Institutes of Health (www.fnih.org). The grantee organization is the Northern California Institute for Research and Education, and the study is coordinated by the Alzheimer's Disease Cooperative Study at the University of California, San Diego. ADNI data are disseminated by the Laboratory for Neuro Imaging at the University of Southern California.





\renewcommand\theequation{\arabic{equation}}
\setcounter{equation}{0}

\renewcommand\thesection{\arabic{section}}
\setcounter{section}{0}

\renewcommand{\thetable}{\arabic{table}}
\setcounter{table}{0}

\renewcommand{\refname}{References for Supplementary Materials}

\section*{\large \bf Web-based Supplementary Materials for ``Multiple Testing for Neuroimaging via Hidden Markov Random Field" by Hai Shu,
	Bin Nan, and
	Robert Koeppe}

\section{Web Appendix A: Interpretations of the Ising Model Parameters}

For the two-parameter Ising model defined in (1) in the main paper, we can show that
\begin{align*}
\label{beta meaning}
\lefteqn{
	\log\Bigg\{\frac{P(\Theta_s=1,\Theta_t=1|\bmath{\theta}_{S\setminus \{s,t\}})}
	{P(\Theta_s=1,\Theta_t=0|\bmath{\theta}_{S\setminus \{s,t\}})}}
\nonumber\\
& \qquad \qquad \times
\frac{P(\Theta_s=0,\Theta_t=0|\bmath{\theta}_{S\setminus \{s,t\}})}{P(\Theta_s=0,\Theta_t=1|\bmath{\theta}_{S\setminus \{s,t\}})}
\Bigg\}
\nonumber\\
& =
\begin{cases}
\beta, & t\in\mathcal{N}(s),\\
0, & \mbox{otherwise.}
\end{cases}
\end{align*}
Therefore, if $s$ and $t$ are neighbors, $\beta$ is equal to a log odds ratio that describes the association between $\Theta_s$ and $\Theta_t$ conditional on all the other state variables being withheld. We can see that $\beta$ reflects how likely the same-state voxels are clustered together. Similarly,
\[
\log \left\{
\frac{P(\Theta_s=1|\sum_{t\in \mathcal{N}(s)}\Theta_t=0)}{P(\Theta_s=0|\bmath\sum_{t\in \mathcal{N}(s)}\Theta_t=0)}
\right\}
=h,
\]
which is the log odds for $\Theta_s=1$ given that $\bmath{\Theta}_{\mathcal{N}(s)}$ are all zero.
Thus, $\beta \ge0$ and $h \le0$ imply the nonnegative dependency of
state variables at neighboring voxels.
In addition, for a voxel $s$ with $m$ nearest neighbors,  we have
\begin{align*}
\label{beta and h meaning}
\lefteqn{\log \Bigg\{\left(
	\frac{P(\Theta_s=1|\sum_{t\in \mathcal{N}(s)}\Theta_t=n)}{P(\Theta_s=0|\bmath\sum_{t\in \mathcal{N}(s)}\Theta_t=n)}\right)}\nonumber\\
&\qquad \qquad \Bigg/
\left(
\frac{P(\Theta_s=0|\sum_{t\in \mathcal{N}(s)}\Theta_t=m-n)}{P(\Theta_s=1|\bmath\sum_{t\in \mathcal{N}(s)}\Theta_t=m-n)}
\right)
\Bigg\}\nonumber\\
&=m\beta+2h,
\end{align*}
where $n$ is an integer satisfying $0\le n \le m$,
which reflects the log ratio of the cluster effect of signals (nonnulls) relative to the cluster effect of noises (nulls).

\section{Web Appendix B: Theoretical Results of the Oracle LIS-Based Procedures for HMRF}
In this section, we show the theoretical results of the oracle LIS-based procedures originally for HMC model in \citet{Sun09} (Theorems 1 to 4 and Corollary 1) and \citet{Wei09} (Theorems 1 and 2), including the validity and optimality of the procedures, also hold for our HMRF model. Here, an FDR procedure is called \emph{valid} if it controls FDR at a prespecified level $\alpha$, and is called \emph{optimal} if it minimizes marginal FNR (mFNR) while controlling marginal FDR (mFDR) at the level $\alpha$.

Unless stated otherwise, the notation in this section is the same as in \citet{Sun09} to which readers are referred. Define $\pi_{ij}=P(\Theta_i=j), i\in S, j=0,1$. The model homogeneity, i.e., $\pi_{ij}=\pi_{j}^{(k)}$ for all $i$ in $k$-th HMC, is required in \citet{Sun09} and in \citet{Wei09} but fails to hold for HMRF because the boundary voxels and interior voxels have different numbers of neighbors.
However, the theory of the oracle procedures still holds for HMRF if we redefine the average conditional cumulative distribution functions (CDFs) of the test statistic $\boldsymbol{T(\boldsymbol{x})}=\{T_i(\boldsymbol{x}): i\in S\}$ by
\begin{equation}\tag{B.1}
\label{B.1}
G^j(t)=\frac{\sum_{i\in S} \pi_{ij} G_i^j(t)}{\sum_{i\in S} \pi_{ij}},
\end{equation}
where $G_i^j(t)=P(T_i<t|\Theta_i=j)$.

For HMC model, \citet{Sun09} proved the optimality of oracle LIS procedure in their Theorems 1 to 3 and Corollary 1, and its validity in their Theorem 4; \citet{Wei09} showed the validity of oracle SLIS procedure in their Theorem 1, and both validity and optimality of oracle PLIS procedure in their Theorem 2.
Let us keep all the statements in these theorems and corollary by
\begin{enumerate}[(i)]
	\item replacing HMM by HMRF;
	\item in Corollary 1 of \citet{Sun09}, replacing the definition of $G^j(t)$ by \eqref{B.1} and the equation $g^1(t)/g^0(t)=(1/t)\pi_0/\pi_1$ by $g^1(t)/g^0(t)=(1/t)\sum_{i \in S}\pi_{i0}/\sum_{i \in S}\pi_{i1}$;
	\item in Theorem 2 of \citet{Wei09}, more precisely stating the optimality of oracle PLIS procedure based on mFDR and mFNR.
\end{enumerate}
For simplicity, we omit all these statements and only provide their proofs in the following.

\subsection{Theorem 1 of \citet{Sun09} for HMRF}
\begin{proof}
	Following the proof of Proposition 1 in \citet{Sun07}, we have
	\begin{equation}
	\label{C.1}
	g^0(c)G^1(c)-G^0(c)g^1(c)>0\tag{C.1}
	\end{equation}
	and
	\begin{equation}
	\label{C.2}
	g^0(c)[1-G^1(c)]-g^1(c)[1-G^0(c)]<0.\tag{C.2}
	\end{equation}
	Additionally, by \eqref{B.1},
	\begin{eqnarray*}
		\mbox{mFDR}(c)&=&\frac{E(N_{10})}{E(R)}=\frac{\sum_{i\in S}P(T_i<c,\Theta_i=0)}{\sum_{i\in S}P(T_i<c)}\\
		&=&\frac{\sum_{i\in S}\pi_{i0}G_i^0(c)}{\sum_{i\in S}(\pi_{i0}G_i^0(c)+\pi_{i1}G_i^1(c))}\\
		&=&\frac{G^0(c)\sum_{i\in S}\pi_{i0}}{G^0(c)\sum_{i\in S}\pi_{i0}+G^1(c)\sum_{i\in S}\pi_{i1}},
	\end{eqnarray*}
	and
	\begin{eqnarray*}
		\mbox{mFNR}(c)&=&\frac{E(N_{01})}{E(S)}=\frac{\sum_{i\in S}P(T_i\ge c,\Theta_i=1)}{\sum_{i\in S}P(T_i\ge c)}\\
		&=&\frac{\sum_{i\in S}\pi_{i1}[1-G_i^1(c)]}{\sum_{i\in S}(\pi_{i0}[1-G_i^0(c)]+\pi_{i1}[1-G_i^1(c)])}\\
		&=&\frac{[1-G^1(c)]\sum_{i\in S}\pi_{i1}}{[1-G^0(c)]\sum_{i\in S}\pi_{i0}+[1-G^1(c)]\sum_{i\in S}\pi_{i1}}.
	\end{eqnarray*}
	Then,
	\begin{align*}
	\lefteqn{\frac{d(\mbox{mFDR}(c))}{dc}}\\
	&=\Bigg(g^0(c)\sum_{i\in S}\pi_{i0}\left[G^0(c)\sum_{i\in S}\pi_{i0}+G^1(c)\sum_{i\in S}\pi_{i1}\right]\\
	&\quad\quad -G^0(c)\sum_{i\in S}\pi_{i0}\left[g^0(c)\sum_{i\in S}\pi_{i0}+g^1(c)\sum_{i\in S}\pi_{i1}\right]\Bigg)\\
	&\quad\quad \Bigg/\left[G^0(c)\sum_{i\in S}\pi_{i0}+G^1(c)\sum_{i\in S}\pi_{i1}\right]^2 \nonumber\\
	&=\frac{[g^0(c)G^1(c)-G^0(c)g^1(c)](\sum_{i\in S}\pi_{i0})(\sum_{i\in S}\pi_{i1})}
	{[G^0(c)\sum_{i\in S}\pi_{i0}+G^1(c)\sum_{i\in S}\pi_{i1}]^2}\\
	&>0
	\end{align*}
	following from \eqref{C.1},
	and
	\begin{align*}
	\lefteqn{\frac{d(\mbox{mFNR}(c))}{dc}}\\
	&=\Bigg\{-g^1(c)\sum_{i\in S}\pi_{i1}\left([1-G^0(c)]\sum_{i\in S}\pi_{i0}+[1-G^1(c)]\sum_{i\in S}\pi_{i1}\right)\\
	&\quad\quad-\left([1-G^1(c)]\sum_{i\in S}\pi_{i1}\right)\left(-g^0(c)\sum_{i\in S}\pi_{i0}-g^1(c)\sum_{i\in S}\pi_{i1}\right)\Bigg\}\notag\\
	&\quad\quad\Bigg/\left([1-G^0(c)]\sum_{i\in S}\pi_{i0}+[1-G^1(c)]\sum_{i\in S}\pi_{i1}\right)^2\nonumber\\
	&=\frac{(g^0(c)[1-G^1(c)]-g^1(c)[1-G^0(c)])(\sum_{i\in S}\pi_{i0})(\sum_{i\in S}\pi_{i1})}{([1-G^0(c)]\sum_{i\in S}\pi_{i0}+[1-G^1(c)]\sum_{i\in S}\pi_{i1})^2}\\
	&<0
	\end{align*}
	following from \eqref{C.2}.
	Hence we obtain part (a) and (b) of the theorem.
	
	For part (c), the classification risk with the loss function
	\[
	L_\lambda(\boldsymbol{\Theta},\boldsymbol{\delta})=\frac{1}{N}\sum_{i \in S}\{
	\lambda(1-\Theta_i)\delta_i+\Theta_i(1-\delta_i)
	\}
	\]
	is
	\begin{align*}
	E[L_\lambda(\boldsymbol{\Theta},\boldsymbol{\delta})]
	&=\frac{1}{N}\sum_{i \in S}\{
	\lambda P(\Theta_i=0,T_i<c)+P(\Theta_i=1,T_i \ge c)\}\\
	&=\frac{1}{N}\sum_{i \in S}\{\lambda\pi_{i0}G_i^0(c)+\pi_{i1}[1-G_i^1(c)]\}\\
	&=\frac{1}{N}\left\{\lambda G^0(c)\sum_{i \in S}\pi_{i0}+[1-G^1(c)]\sum_{i\in S}\pi_{i1}\right\}.
	\end{align*}
	The optimal cutoff $c^*$ that minimizes this risk satisfies
	\[
	\lambda=\frac{g^1(c^*)\sum_{i\in S}\pi_{i1}}{g^0(c^*)\sum_{i \in S}\pi_{i0}}.
	\]
	Since $\boldsymbol{T}\in \mathcal{T}$, we have $g^1(c^*)/g^0(c^*)$ is monotonically decreasing in $c^*$.
	Thus, $\lambda(c^*)$ is monotonically decreasing in $c^*$.
\end{proof}

\subsection{Theorem 2 in \citet{Sun09} for HMRF}
\begin{proof}
	Suppose there are $v_L$ hypotheses from the null and $k_L$ hypotheses from the nonnull among the $r$ rejected hypotheses when the decision rule $\boldsymbol{\delta}(\boldsymbol{L},c_L)$ is applied with test statistic $\boldsymbol{L}$ and cutoff $c_L$. We have
	$v_L=\sum_{i \in S} P(\Theta_i=0,L_i<c_L)$ and $k_L=\sum_{i \in S} P(\Theta_i=1,L_i<c_L)$,
	and
	the classification risk
	\begin{align}
	\label{C.3}
	R_{\lambda(\alpha)}&=E[L_{\lambda(\alpha)}(\boldsymbol{\Theta},\boldsymbol{\delta}(\boldsymbol{L},c_L))]\notag\\
	&=\frac{1}{N}\sum_{i \in S}\{
	\lambda(\alpha) P(\Theta_i=0,L_i<c_L)+P(\Theta_i=1,L_i \ge c_L)\}\notag\\
	&=\frac{1}{N}\left\{\sum_{i \in S}\pi_{i1}+\lambda(\alpha)v_L-k_L\right\}.\tag{C.3}
	\end{align}
	Then following the proof of Theorem 1 in \citet{Sun07} using the expression \eqref{C.3} for the classification risk $R_{\lambda(\alpha)}$, we complete the proof.
\end{proof}

\subsection{Theorems 3 and 4 in \citet{Sun09} for HMRF}
\begin{proof}
	The proofs are the same as those of Theorems 3 and 4 in \citet{Sun09}, thus omitted.
\end{proof}

\subsection{Corollary 1 in \citet{Sun09} for HMRF}
\begin{proof}
	Following the proof of Corollary 1 in \citet{Sun09} with the expression of the risk $R$ replaced by
	\begin{align*}
	R&=\frac{1}{N}\sum_{i \in S}\left\{\frac{1}{t}\pi_{i0}G_i^0(t^*)+\pi_{i1}[1-G_i^1(t^*)]\right\}\\
	&=\frac{1}{N}\left\{\frac{1}{t}G^0(t^*)\sum_{i \in S}\pi_{i0}+[1-G^1(t^*)]\sum_{i \in S}\pi_{i1}\right\}
	\end{align*}
	and their equation $g^1(t^*)/g^0(t^*)=(1/t)\pi_0/\pi_1$ substituted by the new equation $g^1(t^*)/g^0(t^*)=(1/t)\sum_{i \in S}\pi_{i0}/\sum_{i \in S}\pi_{i1}$, we complete the proof.
\end{proof}

\subsection{Theorems 1 and 2 in \citet{Wei09} for HMRF}
\begin{proof}
	For Theorem 1 and the validity of oracle PLIS procedure in Theorem 2, the proofs are the same as those in \citet{Wei09}. For the optimality of oracle PLIS procedure in Theorem 2, the proof is the same as the proof of the optimality
	of oracle LIS procedure given above.
\end{proof}

\section{Web Appendix C: Unbounded likelihood of HMRF}

For any voxel $t\in S$, define a specific configuration of $\bmath{\Theta}$ by $\bmath{\theta}_{\{t\}}=(\theta_s)_{s\in S}$ with $\theta_t=1$ and $\theta_s=0$ if $s\ne t$. Then the observed likelihood function of HMRF
\begin{align*}
L(\bmath{\Phi}|\bmath{x})&=P_{\bmath{\Phi}}(\bmath{x})=
\sum_{\bmath{\Theta}}P_{\bmath{\phi}}(\bmath{x}|\bmath{\Theta})P_{\bmath{\varphi}}(\bmath{\Theta})\\
&\ge P_{\bmath{\phi}}(\bmath{x}|\bmath{\Theta}=\bmath{\theta}_{\{t\}})P_{\bmath{\varphi}}(\bmath{\Theta}=\bmath{\theta}_{\{t\}})\\
&= P_{\bmath{\phi}}(x_t|\Theta_t=1)\prod_{s\in S\setminus \{t\}}P_{\bmath{\phi}}(x_s|\Theta_s=0)P_{\bmath{\varphi}}(\bmath{\Theta}_{S\setminus\{t\}}=\bmath{0},\Theta_t=1)\nonumber\\
&=\left(
\frac{1}{\sqrt{2\pi\sigma_1^2}}\exp\left\{-\frac{(x_t-\mu_1)^2}{2\sigma_1^2}\right\}
+\sum_{l=2}^{L}N(x_t;\mu_l,\sigma^2_l)
\right)\nonumber\\
&\qquad \times
(2\pi)^{-\frac{N-1}{2}}\exp\left\{
-\frac{1}{2}\sum_{s\in S\setminus \{t\}}x_s^2
\right\}
\frac{e^h}{Z(\beta,h)}
\nonumber\\
&\to \infty
\end{align*}
if $\mu_1=x_t$ and $\sigma_1^2\to 0$ with other parameters fixed. Thus the observed likelihood function is unbounded.
The similar unbounded-likelihood phenomenon for
Gaussian hidden Markov chain model has been shown in \citet{Rid97} and \citet{Chen14}.

\section{Web Appendix D: Gibbs Sampler Approximations}
This section presents the approximations of quantities of interest in GEM.
Let $\Omega$ be the set of all possible configurations of $\bmath{\Theta}$:
$
\Omega=\{\bmath{\theta}=(\theta_s)_{s\in S}:\theta_s\in\{0,1\},s\in S\}.
$
By the ergodic theorem of the Gibbs sampler (See Lemma 1 and Theorem 1 in \citet{Robe94}), for any Gibbs distribution (See definition (4.3) in \citet{Gema84}) $\pi(\bmath{\theta})$ and any real-valued function $f(\bmath{\theta})$ on $\Omega$, with probability one,
\[
\lim_{n\to \infty}\frac{1}{n}\sum_{i=1}^nf(\bmath{\theta}^{(i)})=\int_{\Omega}f(\bmath{\theta})d\pi(\bmath{\theta})=E[f(\bmath{\Theta})],
\]
where $\bmath{\theta}^{(i)},i=1,...,n$ are samples successively generated using the Gibbs sampler by $\pi(\bmath{\theta})$.
For our HMRF, it is easy to see that both the Ising model probability distribution $P_{\bmath{\varphi}}(\bmath{\theta})$ and the conditional probability distribution $P_{\bmath{\Phi}^{(t)}}(\bmath{\theta}|\bmath{x})$ are Gibbs distributions. Thus by the ergodic theorem, the following quantities can be approximated using Monte Carlo averages via Gibbs sampler:
\begin{eqnarray*}
	\label{U function}
	\bmath{U}^{(t+1)}({\bmath{\varphi}})
	&=&E_{{\bmath{\Phi}}^{(t)}}[\bmath{H}(\bmath{\Theta})|\bmath{x}]-E_{\bmath{\varphi}}[\bmath{H}(\bmath{\Theta})]\\
	&\approx&\frac{1}{n}\sum_{i=1}^n
	\left(
	\bmath{H}(\bmath{\theta}^{(t,i,\bmath{x})})-\bmath{H}(\bmath{\theta}^{(i,{\bmath{\varphi}})})
	\right), \\
	\label{I function}
	\bmath{I}({\bmath{\varphi}})&=&Var_{\bmath{\varphi}}[\bmath{H}(\bmath{\Theta})]\\
	&=&E_{\bmath{\varphi}}\left[(\bmath{H}(\bmath{\Theta})-E_{\bmath{\varphi}}[\bmath{H}(\bmath{\Theta})])^{\otimes 2}\right]\\
	&\approx&
	\frac{1}{n-1}\sum_{i=1}^n \left(
	\bmath{H}(\bmath{\theta}^{(i, {\bmath{\varphi}})})-\frac{1}{n} \sum_{j=1}^n \bmath{H}(\bmath{\theta}^{(j, {\bmath{\varphi}})})
	\right)^{\bigotimes 2},\\
	\gamma_s^{(t)}(i)&=&P_{{\bmath{\Phi}}^{(t)}}(\Theta_s=i|\bmath{x})=E_{{\bmath{\Phi}}^{(t)}}[\bmath{1}(\Theta_s=i)|\bmath{x}]\\
	&=&E_{{\bmath{\Phi}}^{(t)}}[\bmath{1}(\Theta_s=i)\bmath{1}(\bmath{\Theta}\in\Omega)|\bmath{x}]\\
	&\approx& \frac{1}{n}\sum_{k=1}^n\bmath{1}(\theta_s^{(t,k,\bmath{x})}=i),
\end{eqnarray*}
\begin{eqnarray*}
	\frac{C}{Z({\bmath{\varphi}})}&=&E_{\bmath{\varphi}}[\exp\{-{\bmath{\varphi}}^T\bmath{H}(\bmath{\Theta})\}]\\
	&\approx&\frac{1}{n}\sum_{i=1}^n\exp\{-{\bmath{\varphi}}^T\bmath{H}(\bmath{\theta}^{(i,{\bmath{\varphi}})})\},
\end{eqnarray*}
and
\begin{eqnarray*}
	\lefteqn{Q_2({\bmath{\varphi}}^{(t+1,m)}|{\bmath{\Phi}}^{(t)})-Q_2({\bmath{\varphi}}^{(t)}|{\bmath{\Phi}}^{(t)})}\\
	&=&E_{\bmath{\Phi}^{(t)}}[\log P_{\varphi^{(t+1,m)}}(\bmath{\Theta})-\log P_{\varphi^{(t)}}(\bmath{\Theta})|\bmath{x}]\\
	&=&E_{\bmath{\Phi}^{(t)}}[({\bmath{\varphi}}^{(t+1,m)}-{\bmath{\varphi}}^{(t)})^T\bmath{H}(\bmath{\Theta})|\bmath{x}]+\log\left(\frac{Z(\varphi^{(t)})}{Z(\varphi^{(t+1,m)})}\right)
	\\
	& \approx&
	\frac{1}{n}({\bmath{\varphi}}^{(t+1,m)}-{\bmath{\varphi}}^{(t)})^T\sum_{i=1}^n\bmath{H}(\bmath{\theta}^{(t,i,\bmath{x})})\\
	& &\quad +
	\log \left(
	\frac{\sum_{i=1}^n\exp\{-{{\bmath{\varphi}}^{(t+1,m)}}^T\bmath{H}(\bmath{\theta}^{(i,{\bmath{\varphi}}^{(t+1,m)})})\}}
	{\sum_{i=1}^n\exp\{-{{\bmath{\varphi}}^{(t)}}^T\bmath{H}(\bmath{\theta}^{(i,{\bmath{\varphi}}^{(t)})})\}}
	\right),
\end{eqnarray*}
where $\{\bmath{\theta}^{(1, {\bmath{\varphi}})}, ..., \bmath{\theta}^{(n, {\bmath{\varphi}})}\}$ and $\{\bmath{\theta}^{(t, 1, \bmath{x})}, ..., \bmath{\theta}^{(t, n, \bmath{x})}\}$ are large $n$ samples successively generated using the Gibbs sampler by $P_{\bmath{\varphi}}(\bmath{\theta})$ and $P_{\bmath{\Phi}^{(t)}}(\bmath{\theta}|\bmath{x})$ respectively, and $C$ is the cardinality of set $\Omega$.

\section{Web Appendix E: ADNI FDG-PET Imaging Data Analysis}
Alzheimer's disease (AD) is the most common cause of dementia in the elderly population. The worldwide prevalence of Alzheimer's disease was 26.6 million in 2006 and is predicted to be 1 in 85 persons by 2050 \citep{Broo07}. Much progress has been made in the diagnosis of AD including clinical assessment and neuroimaging techniques. One such extensively used neuroimaging technique is $^{18}\mbox{F}$-Fluorodeoxyglucose positron emission tomography (FDG-PET) imaging, which can be used to evaluate the cerebral metabolic rate of glucose (CMRgl).
Numerous FDG-PET studies
\citep{Nest03,Mosc05b,Lang09}
have demonstrated significant reductions of CMRgl in brain regions in patients with AD and its prodromal stage mild cognitive impairment (MCI), compared with normal control (NC) subjects.
These reduction can be used for the early detection of AD.
Voxel-level multiple testing methods are common approaches to identify
voxels with
significant group differences in CMRgl
\citep{Alex02,Mosc05b,Lang09}. We focus on the comparison between MCI and NC for such a purpose.

The motivating FDG-PET imaging data are obtained from Alzheimer's Disease Neuroimaging Initiative (ADNI) database  ({\tt adni.loni.usc.edu}). These are the baseline FDG-PET images of 102 NC subjects and 206 MCI patients.
Each subjectÕs baseline FDG-PET image has been reoriented into a standard $160 \times 160 \times 96$ voxel image grid with 1.5 mm cubic voxels and the anterior-posterior axis of the subject is parallel to the line connecting the anterior and posterior commissures, so-called AC-PC line. Each image is normalized by the average of voxel values in pons and cerebellar vermis, which are well preserved regions in AlzheimerÕs patients. In human brain, the cerebral cortex is segregated into 43 Brodmann areas (BAs) based on the cytoarchitectural organization of neurons \citep{Gare06}. We consider 30 of them after removing the BAs that are either too small or not always reliably registered. We also investigate 9 subcortical regions, including hippocampus, which are commonly considered in AD studies. A region is further divided into two if its bilateral parts in the left and right hemispheres are separated completely without a shared border in the middle of the brain. We have considered combining neighboring regions to potentially increase accuracy, but failed to find any pair with similar estimated HMRF model parameters. Finally, 61 regions of interest (ROIs) are included in the analysis, where the number of voxels in each region ranges from 149 to 20,680 with a median of 2,517. The total number of voxels of these 61 ROIs is $N = 251,500$.

We apply the PLIS procedure with HMRFs to the analysis of ADNIÕs FDG-PET imaging data, which is compared with BH, $q$-value and CLfdr procedures.
Since the FDG-PET scans were normalized to the average of pons and cerebellar vermis, areas of the brain known to be least affected in AD, it was not surprising that almost all the
signal voxels
are found with decreased
CMRgl.
Both PLIS and CLfdr procedures discovered significant metabolic reduction, with a regional proportion of signals $>$ 50\%, in brain regions preferentially affected by AD, including the posterior cingulate (BAs 23, 31; Mosconi et al., 2008; Langbaum et al., 2009),
parietal cortex (BAs 7, 37, 39, 40; Minoshima et al., 1995; Matsuda, 2001),
temporal cortex (BAs 20 to 22; Alexander et al., 2002; Landau et al., 2011),
medial temporal cortex (BAs 28, 34; Karow et al., 2010),
frontal cortex (BAs 8 to 11, and 44 to 47; Mosconi, 2005),
insular cortex (Perneczky et al., 2007),
amygdala \citep{Nest03}
and hippocampus \citep{Mosc05b}. In regions also typically affected in AD, such as anterior cingulate (BAs 24, 32; Fouquet et al., 2009)
and occipital cortex (BAs 17 to 19; Langbaum et al., 2009),
the proportions of signals found by PLIS are 49.6\% and 39.0\%, respectively, compared with 35.4\% and 11.6\% found by CLfdr,
12.2\% and 0.94\% by $q$-value,
as well as only 1.24\% and 0.87\% by BH.

With respect to the regions that are relatively spared from AD \citep{Bens83,Mats01,Ishi02} or rarely reported in the literature of the disease, caudate, thalamus and putamen are found with high proportions of signals by PLIS ($>$ 45\%) and CLfdr ($>$ 25\%) in each of these regions; signals in medulla, midbrain, cerebellar hemispheres, pre-motor cortex (BA 6)
and primary somatosensory cortex (BAs 1, 2, 3, 5)
are each claimed with a proportion greater than 20\% by PLIS, but very sparse found by
the other three procedures.
Since MCI as a group consists of a mix of patients, many of them will progress to AD but some will not which may include subjects with corticobasal degeneration \citep{Ishi02}, frontotemporal dementia \citep{Jeon05}, or Parkinsonism \citep*{Huan07,Zema11,Ishi13}, it is not surprising that some areas not typical of AD patients were found to be abnormal in the MCI group.

\label{lastpage}


\begin{thebibliography}{}


\bibitem[\protect\citeauthoryear{Benjamini and Hochberg}{1995}]{Benj95}
Benjamini, Y. and Hochberg, Y. (1995). Controlling the false discovery rate: A practical and powerful approach to multiple testing. {\it Journal of the Royal Statistical Society, Series B} {\bf 57,} 289-300.

\bibitem[\protect\citeauthoryear{Benjamini and Hochberg}{2000}]{Benj00}
Benjamini, Y. and Hochberg, Y. (2000). On the adaptive control of the false discovery rate in multiple testing with independent statistics. {\it Journal of Educational and Behavioral Statistics} {\bf 25,} 60-83.

\bibitem[\protect\citeauthoryear{Benjamini and Yekutieli}{2001}]{Benj01}
Benjamini, Y. and Yekutieli, D. (2001). The control of the false discovery rate in multiple testing under dependency. {\it The Annals of Statistics} {\bf 29,} 1165-1188.



\bibitem[\protect\citeauthoryear{Booth and Hobert}{1999}]{Boot99}
Booth, J. G. and Hobert, J. P. (1999). Maximizing generalized linear mixed model likelihoods with an automated Monte Carlo EM algorithm. {\it Journal of the Royal Statistical Society, Series B} {\bf 61,} 265-285.

\bibitem[\protect\citeauthoryear{Bremaud}{1999}]{Brem99}
Bremaud, P. (1999). {\it
Markov Chains: Gibbs Fields, Monte Carlo Simulation, and Queues}. New York: Springer.


\bibitem[\protect\citeauthoryear{Cai and Sun}{2009}]{Cai09}
Cai, T. and Sun, W. (2009). Simultaneous testing of grouped hypotheses: Finding needles in multiple haystacks. {\it Journal of the American Statistical Association} {\bf 104,} 1467-1481.



\bibitem[\protect\citeauthoryear{Chandgotia et~al.}{2014}]{Chan14}
Chandgotia, N., Han, G., Marcus, B., Meyerovitch, T., and Pavlov, R. (2014). One-dimensional Markov random fields, Markov chains and topological Markov fields. {\it Proceedings of the American Mathematical Society} {\bf 142,} 227-242.




\bibitem[\protect\citeauthoryear{Chen, Tan, and Zhang}{Chen et~al.}{2008}]{Chen08}
Chen, J., Tan, X., and Zhang, R. (2008). Inference for normal mixtures in mean and variance. {\it Statistica Sinica} {\bf 18,} 443-465.



\bibitem[\protect\citeauthoryear{Chumbley and Friston}{2009}]{Chum09}
Chumbley, J. R. and Friston, K. J. (2009). False discovery rate revisited: FDR and topological inference using Gaussian random fields. {\it NeuroImage} {\bf 44,} 62-70.

\bibitem[\protect\citeauthoryear{Chumbley et~al.}{2010}]{Chum10}
Chumbley, J., Worsley, K., Flandin, G., and Friston, K. (2010). Topological FDR for neuroimaging. {\it NeuroImage} {\bf 49,} 3057-3064.

\bibitem[\protect\citeauthoryear{Ciuperca, Ridolfi, and Idier}{Ciuperca et~al.}{2003}]{Ciup03}
Ciuperca, G., Ridolfi, A., and Idier, J. (2003). Penalized maximum likelihood estimator for normal mixtures. {\it Scandinavian Journal of Statistics} {\bf 30,} 45-59.



\bibitem[\protect\citeauthoryear{Dabney and Storey}{2014}]{Dabn14}
Dabney, A. and Storey, J. D. (2014). qvalue: Q-value estimation for false discovery rate control. {\it R package version} 1.36.0.


\bibitem[\protect\citeauthoryear{Dempster, Laird, and Rubin}{Dempster et~al.}{1977}]{Demp77}
Dempster, A. P., Laird, N. M., and Rubin, D. B. (1977). Maximum likelihood from incomplete data via the EM algorithm. {\it Journal of the Royal Statistical Society, Series B} {\bf 39,} 1-38.

\bibitem[\protect\citeauthoryear{Efron}{2004}]{Efro04}
Efron, B. (2004). Large-scale simultaneous hypothesis testing: The choice of a null hypothesis. {\it Journal of the American Statistical Association} {\bf 99,} 96-104.



\bibitem[\protect\citeauthoryear{Farcomeni}{2007}]{Farc07}
Farcomeni, A. (2007). Some results on the control of the false discovery rate under dependence. {\it Scandinavian Journal of Statistics} {\bf 34,} 275-297.




\bibitem[\protect\citeauthoryear{Garey}{2006}]{Gare06}
Garey, L. J. (2006). {\it
Brodmann's Localisation in the Cerebral Cortex}. New York: Springer.



\bibitem[\protect\citeauthoryear{Geman and Geman}{1984}]{Gema84}
Geman, S. and Geman, D. (1984). Stochastic relaxation, Gibbs distributions, and the Bayesian restoration of images. {\it IEEE Transactions on Pattern Analysis and Machine Intelligence} {\bf 6,} 721-741.

\bibitem[\protect\citeauthoryear{Genovese, Lazar, and Nichols}{Genovese et~al.}{2002}]{Geno02}
Genovese, C. R., Lazar, N. A., and Nichols, T. (2002). Thresholding of statistical maps in functional neuroimaging using the false discovery rate. {\it NeuroImage} {\bf 15,} 870-878.



\bibitem[\protect\citeauthoryear{Genovese and Wasserman}{2002}]{Geno02b}
Genovese, C. and Wasserman, L. (2002). Operating characteristics and extensions of the false discovery rate procedure. {\it Journal of the Royal Statistical Society, Series B} {\bf 64,} 499-517.

\bibitem[\protect\citeauthoryear{Genovese and Wasserman}{2004}]{Geno04}
Genovese, C. and Wasserman, L. (2004). A stochastic process approach to false discovery control. {\it The Annals of Statistics} {\bf 32,} 1035-1061.


\bibitem[\protect\citeauthoryear{Hoff}{2009}]{Hoff09}
Hoff, P. D. (2009). {\it
A First Course in Bayesian Statistical Methods}. New York: Springer.



\bibitem[\protect\citeauthoryear{Huang et~al.}{2013}]{Huan13}
Huang, L., Goldsmith, J., Reiss, P. T., Reich, D. S., and Crainiceanu, C. M. (2013). Bayesian scalar-on-image regression with application to association between intracranial DTI and cognitive outcomes. {\it NeuroImage} {\bf 83,} 210-223.




\bibitem[\protect\citeauthoryear{Johnson et~al.}{2013}]{John13}
Johnson, T. D., Liu, Z., Bartsch, A. J., and Nichols, T. E. (2013). A Bayesian non-parametric Potts model with application to pre-surgical FMRI data. {\it Statistical Methods in Medical Research} {\bf 22,} 364-381.




\bibitem[\protect\citeauthoryear{Magder and Zeger}{1996}]{Magd96}
Magder, L. S. and Zeger, S. L., (1996). A smooth nonparametric estimate of a mixing distribution using mixtures of Gaussians. {\it Journal of the American Statistical Association} {\bf 91,} 1141-1151.




\bibitem[\protect\citeauthoryear{Nocedal and Wright}{2006}]{Noce06}
Nocedal, J. and Wright, S. (2006). {\it
Numerical Optimization}, 2nd edition. New York: Springer.


\bibitem[\protect\citeauthoryear{Ridolfi}{1997}]{Rido97}
Ridolfi, A. (1997). Maximum likelihood estimation of hidden Markov model parameters, with application to medical image segmentation. Politecnico di Milano, Milan, Italy.

\bibitem[\protect\citeauthoryear{Roberts and Smith}{1994}]{Robe94}
Roberts, G. O. and Smith A. F. M. (1994). Simple conditions for the convergence of the Gibbs sampler and Metropolis-Hastings algorithms. {\it Stochastic Processes and their Applications} {\bf 49,} 207-216.

\bibitem[\protect\citeauthoryear{Stoer and Bulirsch}{2002}]{Stoe02}
Stoer, J. and Bulirsch, R. (2002). {\it
Introduction to Numerical Analysis}, 3rd edition. New York: Springer.

\bibitem[\protect\citeauthoryear{Storey}{2003}]{Store03}
Storey, J. D. (2003). The positive false discovery rate: A Bayesian interpretation and the $q$-value. {\it The Annals of Statistics} {\bf 31,} 2013-2035.

\bibitem[\protect\citeauthoryear{Sun and Cai}{2007}]{Sun07}
Sun, W. and Cai, T. T. (2007). Oracle and adaptive compound decision rules for false discovery rate control. {\it Journal of the American Statistical Association} {\bf 102,} 901-912.

\bibitem[\protect\citeauthoryear{Sun and Cai}{2009}]{Sun09}
Sun, W. and Cai, T. T. (2009). Large-scale multiple testing under dependence. {\it Journal of the Royal Statistical Society, Series B} {\bf 71,} 393-424.

\bibitem[\protect\citeauthoryear{Wei et~al.}{2009}]{Wei09}
Wei, Z., Sun, W., Wang, K., and Hakonarson, H. (2009). Multiple testing in genome-wide association studies via hidden Markov models. {\it Bioinformatics} {\bf 25,} 2802-2808.

\bibitem[\protect\citeauthoryear{Welch}{1947}]{Welc47}
Welch, B. L. (1947). The generalization of `Student's' problem when several different population variances are involved. {\it Biometrika} {\bf 34,} 28-35.

\bibitem[\protect\citeauthoryear{Winkler}{2003}]{Wink03}
Winkler, G. (2003). {\it
Image Analysis, Random Fields and Markov Chain Monte Carlo Methods}, 2nd edition. New York: Springer.




\bibitem[\protect\citeauthoryear{Wu}{2008}]{Wu08}
Wu, W. B. (2008). On false discovery control under dependence. {\it The Annals of Statistics} {\bf 36,} 364-380.



\bibitem[\protect\citeauthoryear{Zhang, Fan, and Yu}{2011}]{Zhan11}
Zhang, C, Fan, J., and Yu, T. (2011). Multiple testing via FDR\_L for large-scale image data. {\it The Annals of Statistics} {\bf 39,} 613-642.


\bibitem[\protect\citeauthoryear{Zhang et~al.}{2008}]{Zhan08}
Zhang, X., Johnson, T. D., Little, R. J. A., and Cao, Y. (2008). Quantitative magnetic resonance image analysis via the EM algorithm with stochastic variation. {\it The Annals of Applied Statistics} {\bf 2,} 736-755.



\end{thebibliography}

\begin{thebibliography}{}

	\bibitem[\protect\citeauthoryear{Alexander et~al.}{2002}]{Alex02}
	Alexander, G. E., Chen, K., Pietrini, P., Rapoport, S. I., and Reiman, E. M. (2002). Longitudinal PET evaluation of cerebral metabolic decline in dementia: A potential outcome measure in Alzheimer's disease treatment studies. {\it American Journal of Psychiatry} {\bf 196,} 738-745.
	
	\bibitem[\protect\citeauthoryear{Benson et~al.}{1983}]{Bens83}
	Benson, D. F., Kuhl, D. E., Hawkins, R. A., Phelps, M. E., Cummings, J. L., and Tsai, S. Y. (1983). The fluorodeoxyglucose 18F scan in Alzheimer's disease and multi-infarct dementia. {\it Archives of Neurology} {\bf 40,} 711-714.
	
	\bibitem[\protect\citeauthoryear{Brookmeyer et~al.}{2007}]{Broo07}
	Brookmeyer, R., Johnson, E., Ziegler-Graham, K., and Arrighi, H. M. (2007). Forecasting the global burden of AlzheimerÕs disease. {\it Alzheimer's \& Dementia} {\bf 3,} 186-191.
	
	\bibitem[\protect\citeauthoryear{Chen, Huang, and Wang}{2014}]{Chen14}
	Chen, J., Huang, Y., and Wang, P. (2014). Composite likelihood under hidden Markov model. {\it Statistica Sinica} [Preprint], doi:10.5705/ss.2013.084t.
	
	\bibitem[\protect\citeauthoryear{Fouquet et~al.}{2009}]{Fouq09}
	Fouquet, M., Desgranges, B., Landeau, B., Duchesnay, E., Mezenge, F., De La Sayette, V., et al. (2009). Longitudinal brain metabolic changes from amnestic mild cognitive impairment to Alzheimer's disease. {\it Brain} {\bf 132,} 2058-2067.
	
	\bibitem[\protect\citeauthoryear{Garey}{2006}]{Gare06}
	Garey, L. J. (2006). {\it
		Brodmann's Localisation in the Cerebral Cortex}. New York: Springer.
	
	\bibitem[\protect\citeauthoryear{Geman and Geman}{1984}]{Gema84}
	Geman, S. and Geman, D. (1984). Stochastic relaxation, Gibbs distributions, and the Bayesian restoration of images. {\it IEEE Transactions on Pattern Analysis and Machine Intelligence} {\bf 6,} 721-741.
	
	
	\bibitem[\protect\citeauthoryear{Huang et~al.}{2007}]{Huan07}
	Huang, C., Tang, C., Feigin, A., Lesser, M., Ma, Y., Pourfar, M., et al. (2007). Changes in network activity with the progression of ParkinsonÕs disease. {\it Brain} {\bf 130,} 1834-1846.
	
	\bibitem[\protect\citeauthoryear{Ishii}{2002}]{Ishi02}
	Ishii, K. (2002). Clinical application of positron emission tomography
	for diagnosis of dementia. {\it Annals of Nuclear Medicine} {\bf 16,} 515-525.
	
	
	\bibitem[\protect\citeauthoryear{Ishii}{2013}]{Ishi13}
	Ishii, K. (2013). PET approaches for diagnosis of dementia. {\it American Journal of Neuroradiology} [online], DOI: 10.3174/ajnr.A3695.
	
	\bibitem[\protect\citeauthoryear{Jeong et~al.}{2005}]{Jeon05}
	Jeong, Y., Cho, S. S., Park, J. M., Kang, S. J., Lee, J. S., Kang, E., et al. (2005). 18F-FDG PET findings in frontotemporal dementia: An SPM analysis of 29 patients. {\it Journal of Nuclear Medicine} {\bf 46,} 233-239.
	
	\bibitem[\protect\citeauthoryear{Karow et~al.}{2010}]{Karo10}
	Karow, D. S., McEvoy, L. K., Fennema-Notestine, C., Hagler, D. J., Jennings, R. G., Brewer, J. B., et al. (2010).
	Relative capability of MR imaging and FDG PET to depict changes associated with prodromal and early Alzheimer disease. {\it Radiology} {\bf 256,} 932-942.
	
	\bibitem[\protect\citeauthoryear{Landau et~al.}{2011}]{Land11}
	Landau, S. M., Harvey, D., Madison, C. M., Koeppe, R. A., Reiman, E. M., Foster, N. L., et al. (2011). Associations between cognitive, functional, and FDG-PET measures of decline in AD and MCI. {\it Neurobiology of Aging} {\bf 32,} 1207-1218.
	
	\bibitem[\protect\citeauthoryear{Langbaum et~al.}{2009}]{Lang09}
	Langbaum, J. B. S., Chen, K.,  Lee, W., Reschke, C., Bandy, D., Fleisher, A. S., et al. (2009). Categorical and correlational analyses of baseline fluorodeoxyglucose positron emission tomography images from the Alzheimer's Disease Neuroimaging Initiative (ADNI). {\it NeuroImage} {\bf 45,} 1107-1116.
	
	\bibitem[\protect\citeauthoryear{Matsuda}{2001}]{Mats01}
	Matsuda, H. (2001). Cerebral blood flow and metabolic abnormalities in AlzheimerÕs disease. {\it Annals of Nuclear Medicine} {\bf 15,} 85-92.
	
	\bibitem[\protect\citeauthoryear{Minoshima et~al.}{1995}]{Mino95}
	Minoshima, S., Frey, K. A., Koeppe, R. A., Foster, N. L., and Kuhl, D. E. (1995). A diagnostic approach in Alzheimer's disease using three-dimensional stereotactic surface projections of Fluorine-18-FDG PET. {\it Journal of Nuclear Medicine} {\bf 36,} 1238-1248.
	
	
	\bibitem[\protect\citeauthoryear{Mosconi}{2005}]{Mosc05}
	Mosconi, L. (2005). Brain glucose metabolism in the early and specific diagnosis of Alzheimer's disease. {\it European Journal of Nuclear Medicine and Molecular Imaging} {\bf 32,} 486-510.
	
	\bibitem[\protect\citeauthoryear{Mosconi et~al.}{2005}]{Mosc05b}
	Mosconi, L., Tsui, W. H., De Santi, S., Li, J., Rusinek, H., Convit, A., et al. (2005). Reduced hippocampal metabolism in MCI and AD: Automated FDG-PET image analysis. {\it Neurology} {\bf 64,} 1860-1867.
	
	\bibitem[\protect\citeauthoryear{Mosconi et~al.}{2008}]{Mosc08}
	Mosconi, L., Tsui, W. H., Herholz, K., Pupi, A., Drzezga, A., Lucignani, G., et al. (2008). Multicenter standardized 18F-FDG PET diagnosis of mild cognitive impairment, Alzheimer's disease, and other dementias. {\it Journal of Nuclear Medicine} {\bf 49,} 390-398.
	
	
	\bibitem[\protect\citeauthoryear{Nestor et~al.}{2003}]{Nest03}
	Nestor, P. J., Fryer, T. D., Smielewski, P., and Hodges, J. R. (2003). Limbic hypometabolism in Alzheimer's disease and mild cognitive impairment. {\it Annals of Neurology} {\bf 54,} 343-351.
	
	\bibitem[\protect\citeauthoryear{Perneczky et~al.}{2007}]{Pern07}
	Perneczky, R., Drzezga, A., Diehl-Schmid, J., Li, Y., and Kurz, A. (2007). Gender differences in brain reserve: An (18)F-FDG PET study in AlzheimerÕs disease. {\it Journal of Neurology} {\bf 254,} 1395-1400.
	
	\bibitem[\protect\citeauthoryear{Ridolfi}{1997}]{Rid97}
	Ridolfi, A. (1997). Maximum likelihood estimation of hidden Markov model parameters, with
	application to medical image segmentation. Politecnico di Milano, Milan, Italy.
	
	\bibitem[\protect\citeauthoryear{Roberts and Smith}{1994}]{Robe94}
	Roberts, G. O. and Smith A. F. M. (1994). Simple conditions for the convergence of the Gibbs sampler and Metropolis-Hastings algorithms. {\it Stochastic Processes and their Applications} {\bf 49,} 207-216.
	
	\bibitem[\protect\citeauthoryear{Sun and Cai}{2007}]{Sun07}
	Sun, W. and Cai, T. T. (2007). Oracle and adaptive compound decision rules for false discovery rate control. {\it Journal of the American Statistical Association} {\bf 102,} 901-912.
	
	\bibitem[\protect\citeauthoryear{Sun and Cai}{2009}]{Sun09}
	Sun, W. and Cai, T. T. (2009). Large-scale multiple testing under dependence. {\it Journal of the Royal Statistical Society, Series B} {\bf 71,} 393-424.
	
	\bibitem[\protect\citeauthoryear{Wei et~al.}{2009}]{Wei09}
	Wei, Z., Sun, W., Wang, K., and Hakonarson, H. (2009). Multiple testing in genome-wide association studies via hidden Markov models. {\it Bioinformatics} {\bf 25,} 2802-2808.
	
	
	\bibitem[\protect\citeauthoryear{Zeman, Carpenter, and Scott}{Zeman et~al.}{2011}]{Zema11}
	Zeman, M. N., Carpenter, G. M., and Scott, P. J. (2011). Diagnosis of dementia using nuclear medicine imaging modalities. In {\it 12 Chapters on Nuclear Medicine}. Croatia: InTech, pp. 199-230.
	
\end{thebibliography}
\end{document}